\documentclass{article}

\PassOptionsToPackage{numbers, compress}{natbib}
\usepackage[preprint]{neurips_2026}

\usepackage[utf8]{inputenc} % allow utf-8 input
\usepackage[T1]{fontenc}    % use 8-bit T1 fonts
\usepackage{hyperref}       % hyperlinks
\usepackage{url}            % simple URL typesetting
\usepackage{booktabs}       % professional-quality tables
\usepackage{graphicx}       % resizebox
\usepackage{multirow}       % merged table rows
\usepackage{enumitem}       % compact list options such as noitemsep
\usepackage{amsmath}        % align environment and math formatting
\usepackage{amsfonts}       % blackboard math symbols
\usepackage{amssymb}        % symbols such as \triangleq and \varnothing
\usepackage{caption}       % \captionof inside wrapfigure (half-column pseudocode)
\usepackage{wrapfig}        % wrap Algorithm beside prose (half width)
\usepackage{nicefrac}       % compact symbols for 1/2, etc.
\usepackage{microtype}      % microtypography
\usepackage{xcolor}         % colors
\usepackage{fvextra}        % improved verbatim blocks
\usepackage{algpseudocodex}
\usepackage[ruled, lined, commentsnumbered]{algorithm2e}

\DefineVerbatimEnvironment{PromptBlock}{Verbatim}{
  breaklines=true,
  breakanywhere=true,
  fontsize=\footnotesize,
  frame=single,
  framerule=0.4pt,
  rulecolor=\color{black!18},
  framesep=4mm,
  xleftmargin=1em,
  xrightmargin=1em,
  baselinestretch=0.95,
  formatcom=\color{black!92}
}

\title{Enhancing LLMs with Cognitive-Affective \\Personality Inference for Simulating \\Human Social-Psychological Behavior}

\author{%
  \normalfont
  \textbf{Zhibo Deng}\textsuperscript{1}
  \quad
  \textbf{Dongyuan Li}\textsuperscript{1}\thanks{%
    Corresponding to: Dongyuan Li and Renhe Jiang.
  }
  \quad
  \textbf{Shuwen Ge}\textsuperscript{1}
  \quad
  \textbf{Ziqing Zhang}\textsuperscript{1}
  \\
  \textbf{Ying Zhang}\textsuperscript{2}
  \quad
  \textbf{Renhe Jiang}\textsuperscript{1}\footnotemark[1]
  \\
  \normalfont
  \textsuperscript{1}The University of Tokyo
  \\
  \textsuperscript{2}RIKEN Center for Advanced Intelligence Project
  \\
  {
    \texttt{danel0727@outlook.com}
    \quad
    \texttt{shuwen8681@gmail.com}
  }
  \\
  {
    \texttt{ying.zhang@riken.jp}
    \quad
    \texttt{zhangzq09@gmail.com}
  }
  \\
  {
    \texttt{\{lidy,jiangrh\}@csis.u-tokyo.ac.jp}
  }
}

\begin{document}

\makeatletter
\renewcommand{\@fnsymbol}[1]{\ensuremath{\dagger}}
\maketitle
\makeatother

\begin{abstract}
Large language models are increasingly used to simulate human participants in social and behavioral studies, yet static persona prompting typically maps a participant profile and an experimental scenario directly to a response, entangling stable dispositions with situation-specific interpretations. To address this limitation, we introduce \textbf{SPIN}, a cognitive-affective personality system-inspired inference pipeline for simulating human social-psychological behavior. Specifically, SPIN implements this structured inference process through three zero-shot LLM calls that compile a task-blind participant core, elicit condition-specific cognitive-affective states, and read out decisions from those states, thereby reusing stable personality structure while routing each trial-specific response through an explicit state representation. We evaluate SPIN on two reconstructed social-psychological study families spanning uncertainty reasoning and pluralistic ignorance, across four base LLMs. Compared with blank, demographic, narrative, and chain-of-thought prompt variants, SPIN consistently delivers the strongest overall alignment performance across base LLMs and study families. Ablations and state analyses further show that both personality compilation and structured state elicitation contribute to the gains, and that the elicited states shift interpretably across informational and normative conditions. These results suggest that structured personality-state inference can improve benchmark-level behavioral alignment beyond richer persona descriptions or generic multi-step reasoning. Our code is publicly available at \url{https://github.com/DDanel0727/spinn}.
\end{abstract}

\section{Introduction}
Large language models (LLMs) are increasingly being used as synthetic respondents in social and behavioral research. Conditioned on demographic attributes, attitudinal profiles, or richer life-history descriptions, they can approximate aggregate response patterns in surveys, economic games, and classic behavioral experiments~\citep{horton2023large,aher2023using,argyle2023out,park2024generative,binz2025foundation}. Recent large-scale replications further show strong correlations between simulated and actual treatment effects across dozens to hundreds of studies~\citep{hewitt2024predicting,cui2025large}. 
However, for scalable pilot studies and preliminary theory testing, surface plausibility is insufficient: model-generated behavior should align with the human behavioral summaries that social-science studies are designed to measure.

Existing evidence shows that this alignment remains fragile. LLM-generated samples may compress human variability, miscalibrate effect magnitudes, and miss subtle cross-condition contrasts~\citep{gui2023challenge,gao2025take,hullman2026human,bisbee2024synthetic}; even when aggregate means are well matched, response distributions can exhibit lower variance and reduced demographic sensitivity than real survey data~\citep{anthis2025llm}. Individual responses can also drift from the assigned personas~\citep{abdulhai2025consistently}. A central modeling limitation lies in end-to-end prompting: whether the prompt contains no profile, demographics, or a detailed narrative, the model usually maps the available participant description and scenario directly to an answer. Stable dispositions and situation-specific interpretations are then inferred in one generation step, making it difficult to reuse a participant consistently while remaining sensitive to small manipulations.

We operationalize simulation quality as benchmark-level behavioral alignment: matching simulated responses to theory-relevant human behavioral summaries---including choice rates, rating statistics, and cross-condition contrasts---rather than recovering full individual psychologies. To this end, we introduce \textbf{SPIN} (\textbf{S}tructured \textbf{P}ersonality-\textbf{I}nformed \textbf{IN}ference), a zero-shot staged inference pipeline inspired by the Cognitive-Affective Personality System (CAPS) account of stable dispositions and situation-contingent behavior~\citep{caps-1995,mischel2008toward}. SPIN uses three LLM calls to compile a task-blind personality core, elicit a condition-specific cognitive-affective state, and read out the benchmark-facing decision from the elicited state. Importantly, stagewise input restrictions prevent personality compilation from seeing the trial, prevent state elicitation from accessing the raw profile, and ground decision readout in the elicited state. The compiled core is reused across trials, while states are recomputed for each trial or item-level condition.

We evaluate SPIN on two reconstructed social-psychological study families: Shafir and Tversky's uncertainty-reasoning paradigms~\citep{shafir1992thinking} and Prentice and Miller's pluralistic-ignorance experiments~\citep{prentice1993pluralistic}. Across four LLMs, SPIN achieves the strongest benchmark-level alignment among blank prompting, demographic conditioning, narrative backstories, and chain-of-thought variants. Ablations show that both task-blind personality compilation and condition-specific state elicitation matter, and aggregate state analyses show interpretable shifts across informational and normative conditions. These results support the narrower claim that structured personality--state inference can improve benchmark-level behavioral alignment beyond richer persona descriptions or generic multi-step reasoning.

Overall, our main contributions can be summarized as follows:
\begin{itemize}[leftmargin=*, itemsep=1.2pt]
  \item We formalize social-psychological behavior alignment as benchmark-level matching of theory-relevant human behavioral summaries, including choice rates, rating statistics, and cross-condition contrasts, and identify a core modeling limitation of end-to-end prompting: stable participant dispositions and condition-specific interpretations are entangled within a single generation step.
% Section 3

  \item We introduce SPIN, a zero-shot CAPS-inspired staged inference pipeline that separates response generation into task-blind personality compilation, condition-specific cognitive-affective state elicitation, and state-grounded decision readout. SPIN uses structured intermediate representations and stagewise input restrictions to route trial-specific responses through an explicit personality--state bottleneck.
% Section 4

  \item We evaluate SPIN on two reconstructed social-psychological study families spanning uncertainty reasoning and pluralistic ignorance across four base LLMs. SPIN achieves the strongest benchmark-level alignment over blank, demographic, narrative, and their chain-of-thought augmented variants; ablations and aggregate state analyses further provide diagnostic evidence that both personality compilation and structured state elicitation contribute to these gains.
% Section 5
\end{itemize}

\section{Related Work}
LLMs have been widely studied as synthetic respondents for surveys, economic games, and behavioral experiments, with evidence that demographic, attitudinal, or biographical conditioning can reproduce some aggregate human response patterns~\citep{horton2023large,aher2023using,argyle2023out,park2024generative,hewitt2024predicting,cui2025large}. At the same time, validation work shows that such simulations can compress human variability, miscalibrate effects, miss subgroup or cross-condition structure, and drift from assigned personas; recent benchmarks therefore emphasize held-out human outcomes and normalized behavioral summaries rather than raw generations alone~\citep{gui2023challenge,bisbee2024synthetic,gao2025take,hullman2026human,anthis2025llm,abdulhai2025consistently,liu2026humanstudy}. Existing remedies mainly enrich the persona or add generic reasoning steps, including narrative backstories, character-consistency methods, chain-of-thought, and context-navigation strategies~\citep{jiang2024personallm,du2025simvbg,wei2022chain,kojima2022large,kong2026improving}; however, these do not by themselves ensure consistent participant reuse across trials or isolate the situation-specific state that mediates a decision. A complementary line uses cognitive and psychological theory to analyze or structure LLM behavior, from cognitive benchmarks and value-tradeoff models to cognitively faithful decision processes, belief elicitation, and executable models of others' routines~\citep{coda2024cogbench,murthycognitive,cousins2025towards,arghal2025steering,jha2025modeling}. This work builds on that direction by using CAPS~\citep{caps-1995,mischel2008toward} to separate stable dispositions from situation-specific activation for benchmark-level social judgment alignment. A fuller discussion is provided in Appendix~\ref{app:related}.

\section{Task Formulation}
\label{sec:task}

Let $\mathcal{P}=\{p_1,\dots,p_N\}$ be a set of participants, each associated with a profile $x_p$ containing demographic, behavioral, or personality information, and let $\mathcal{T}=\{\tau_1,\dots,\tau_M\}$ be a set of trials, each specifying a participant-facing scenario, an experimental condition, a response space, and a measurement protocol. For each profile–trial pair $(x_p,\tau)$, a model generates a benchmark-compatible response, which the benchmark evaluator maps into study-level behavioral summaries such as response proportions, rating statistics, and cross-condition contrasts. We define benchmark-level behavioral alignment as matching these simulated summaries to the corresponding human targets, rather than recovering each participant's full latent psychology or complete individual response trajectory. The prevailing approach implements response generation as a single LLM inference call:
\begin{equation}
  \label{eq:e2e}
  \hat{Y}_{p,\tau} \;\sim\; f_{\textsc{llm}}(x_p,\;\tau),
\end{equation}
where the participant profile and the trial stimulus are concatenated into a monolithic prompt. Although straightforward, this formulation asks the model to infer stable participant tendencies and condition-specific interpretations within the same generation step, so the same profile may be reused inconsistently across trials and subtle condition changes may be under- or over-weighted in the final response. SPIN addresses this by separating stable personality compilation, condition-specific state elicitation, and decision readout. We introduce these stages in detail in the following subsections.

\section{Methodology}
\label{sec:method}

We present SPIN, a structured, personality-informed inference pipeline for social-psychological behavior alignment that decomposes the generation of behavioral responses into three sequential stages, inspired by the CAPS theory \citep{caps-1995,mischel2008toward}. Rather than inferring behavior in a single pass from a raw participant profile and task description, SPIN first compiles stable personality characteristics, then activates a transient cognitive-affective state conditioned on the current situation, and finally produces a benchmark-facing decision grounded in that state. In the remainder of this section, we first illustrate the overall architecture of SPIN (Section~\ref{sec:overview}), followed by a detailed breakdown of its three inference stages (Sections~\ref{sec:stage1}--\ref{sec:stage3}) and the structural design principles intended to improve behavioral alignment (Section~\ref{sec:principles}).

\subsection{Framework Overview: A CAPS-Inspired Architecture}
\label{sec:overview}

Among cognitive theories of personality and behavior, CAPS offers a particularly useful account of why human behavior is simultaneously stable and variable: it posits a network of stable cognitive-affective units---encodings, expectancies, goals, values, and self-regulatory plans---that are selectively activated by situational features, producing predictable if–then patterns of situation-behavior variation rather than uniform responses. This distinction between stable dispositions and transient situational activations directly addresses the two limitations of end-to-end prompting identified above, explaining how both contribute to observable behavior through a structured mediating process. We instantiate this distinction with a small fixed vocabulary of dispositional traits and situational state variables, designed not as a direct import from a single psychometric inventory but as a compact modeling interface covering recurrent axes of social judgment and strategic interaction. Appendix~\ref{app:dimension_justification} summarizes the dimension-level rationale.

\begin{figure*}[t]
  \centering
  \includegraphics[width=1\textwidth]{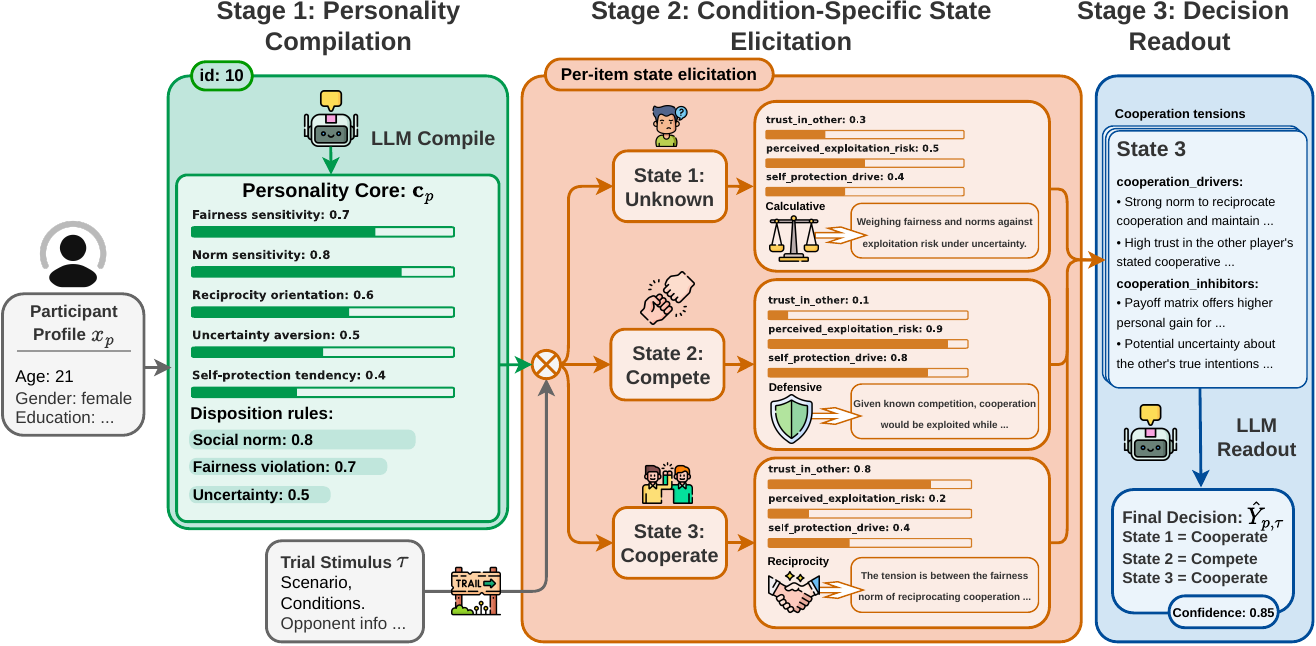}
  \caption{Overview of the SPIN pipeline for participant-aligned decision generation. SPIN decomposes behavioral generation into three sequential stages. Stage 1 compiles a participant profile $x_p$ into a stable personality core $\mathbf{c}_p$ consisting of trait scores and disposition rules; this core is computed once per participant and reused across subsequent trials. Stage 2 takes $\mathbf{c}_p$ and a trial stimulus $\tau$ as input and performs item-wise state elicitation, so that each experimental condition yields its own transient state. Stage 3 converts the resulting indexed collection of elicited states into a response $\hat{Y}_{p,\tau}$, producing per-item choices and an overall confidence score.}
  \label{fig:pipeline}
\end{figure*}

We operationalize this insight as a three-stage inference pipeline (illustrated in Figure~\ref{fig:pipeline}). Rather than mapping a profile--trial pair directly to a response in a single pass, SPIN is designed to first \textsc{Compile} stable personality characteristics from the participant profile, then \textsc{Elicit} a transient cognitive-affective state conditioned on the current situation, and finally \textsc{Decide} by reading out a behavioral response grounded in that state:
\begin{equation}
  \label{eq:decomposition}
  \hat{Y}_{p,\tau} \;\sim\;
    \textsc{Decide}\!\Big(
      \textsc{Elicit}\!\big(
        \textsc{Compile}(x_p),\;\tau
      \big),\;\tau
    \Big).
\end{equation}
Before producing benchmark-facing responses, SPIN constructs structured intermediates---a reusable personality core and trial-specific elicited states---making it personality-informed rather than merely profile-conditioned. Operationally, the three operators are implemented as stage-specific prompt templates applied to the same backbone LLM (not separate trained models) and each is executed as an independent LLM call with a fresh context window containing only the inputs specified for that stage, with no carry-over of previous-stage raw inputs. Stage outputs follow a declared JSON schema; completions are parsed, and trait/state scalars are clipped to $[0,1]$ where specified. \textsc{Compile} is called once per participant and cached; \textsc{Elicit} is called once per trial, or once per item when distinct item-level conditions are present; and \textsc{Decide} is called once per trial to produce per-item choices and the canonical answer line. All prompts are zero-shot with temperature $1.0$, a fixed decoding configuration, and no self-consistency voting or multi-sample aggregation; representative templates are provided in Appendix~\ref{app:prompts}.

\subsection{Stage 1: Personality Compilation}
\label{sec:stage1}

The first stage compiles the participant profile $x_p$ into a structured personality core $\mathbf{c}$:
\begin{equation}
  \label{eq:core}
  \mathbf{c}_p
  = \bigl(\,\boldsymbol{\theta},\;\mathcal{R},\;\delta\,\bigr)
  = \textsc{Compile}(x_p),
\end{equation}
consisting of \textbf{(i)}~a continuous trait vector $\boldsymbol{\theta}\in[0,1]^K$ capturing $K$ dispositional dimensions relevant to social decision tasks (in our implementation, $K\!=\!5$: fairness sensitivity, norm sensitivity, reciprocity orientation, uncertainty aversion, and self-protection tendency); \textbf{(ii)}~a disposition rule set $\mathcal{R}=\{(\text{cue}_j,\text{tendency}_j,\text{strength}_j)\}_{j=1}^{J}$ encoding cue-triggered behavioral tendencies. The number of rules is participant-specific rather than fixed; each rule pairs a cue from a fixed cue vocabulary with a short natural-language tendency and a continuous strength in $[0,1]$, e.g., betrayal risk $\rightarrow$ reduces cooperation under uncertainty with strength $0.7$; and \textbf{(iii)}~a default decision-style summary $\delta$ in natural language. This stage is task-blind: it receives only $x_p$ and has no access to any trial information, ensuring that the resulting core captures stable participant-level structure rather than ad hoc trial-specific reasoning.

\subsection{Stage 2: Condition-Specific State Elicitation}
\label{sec:stage2}

The second stage maps the personality core and the current trial stimulus to a transient cognitive-affective state:
\begin{equation}
  \label{eq:state}
  \mathbf{s}_{p,\tau}^{(i)}
  = \textsc{Elicit}(\mathbf{c}_p,\;\tau),
\end{equation}
where the index $i$ denotes the item-level condition being elicited within the trial $\tau$. For single-item trials, there is only one state; while for multi-item trials, we instantiate the same elicitation interface separately for each item or condition, yielding an indexed collection of states. Each elicited state contains a continuous state vector $\boldsymbol{\sigma}\in[0,1]^L$ ($L\!=\!6$: arousal, conflict, social concern, trust, perceived exploitation risk, and self-protection drive), response-promoting and response-inhibiting considerations, a dominant decision mode selected from a fixed set (e.g., deliberative, cautious, defensive), and a short rationale linking the personality core to the current condition. This stage receives the compiled core $\mathbf{c}_p$ but not the raw profile $x_p$, so all personality information entering the state must pass through the structured core. Item-wise elicitation is used when a trial contains distinct experimental conditions, allowing the same personality core to produce condition-sensitive states without blurring the variation the benchmark is designed to test.

\subsection{Stage 3: Decision Readout}
\label{sec:stage3}

The final stage collects the item-indexed state elicited in Stage~2 and converts them into a response. We write this collection as
$\mathcal{S}_{p,\tau}=(\mathbf{s}_{p,\tau}^{(i)})_{i=1}^{n_\tau}$, where the index preserves the correspondence between each item-level condition and its elicited state:
\begin{equation}
  \label{eq:decision}
  \hat{Y}_{p,\tau}
  = \textsc{Decide}(\mathcal{S}_{p,\tau},\;\tau).
\end{equation}
The stage produces per-item choices, a response confidence score, and a canonical answer line in the format required by the benchmark. For multi-item trials, each item-level decision is grounded in its corresponding indexed state rather than blended across items. This stage receives neither the raw profile nor the personality core. The decision is grounded entirely in $\mathcal{S}_{p,\tau}$.

\begin{wrapfigure}{r}{0.5\textwidth}
  \centering
  \vspace{-1.5em}
  \begin{minipage}{\linewidth}
    \footnotesize
\begin{algorithm}[H]
  \SetAlgoLined
  \SetAlFnt{\footnotesize}
  \SetAlCapFnt{\footnotesize}
  \SetAlCapNameFnt{\footnotesize}

  \SetKwInOut{Input}{Input}
  \SetKwInOut{Output}{Output}

  \caption{The SPIN inference procedure.}
  \label{alg:spin-inference}

  \Input{participant profile $x_p$; trial set $\mathcal{T}$}
  \Output{response $\hat{Y}_{p,\tau}$ for each $\tau\in\mathcal{T}$}

  $\mathbf{c}_p \gets \textsc{Compile}(x_p)$\;

  \For{each trial $\tau \in \mathcal{T}$}{
    \eIf{$\tau$ has item-level conditions $\{q_1,\dots,q_n\}$}{
      \For{each item $q_i$}{
        $\mathbf{s}_{p,\tau}^{(i)} \gets \textsc{Elicit}(\mathbf{c}_p, q_i)$\;
      }
      $\mathcal{S}_{p,\tau} \gets (\mathbf{s}_{p,\tau}^{(i)})_{i=1}^{n}$\;
    }{
      $\mathbf{s}_{p,\tau}^{(1)} \gets \textsc{Elicit}(\mathbf{c}_p,\tau)$\;
      $\mathcal{S}_{p,\tau} \gets (\mathbf{s}_{p,\tau}^{(1)})$\;
    }
    $\hat{Y}_{p,\tau} \gets \textsc{Decide}(\mathcal{S}_{p,\tau},\tau)$\;
  }
\end{algorithm}
  \end{minipage}
  \vspace{-1.5em}
\end{wrapfigure}

\subsection{Structural Design Principles}
\label{sec:principles}

Algorithm~\ref{alg:spin-inference} summarizes the inference procedure. Several design choices shape how this decomposition works in practice. The compilation stage receives only the participant profile and is entirely task-blind, so the resulting personality core is constructed without access to trial-specific information. The elicitation stage then receives the compiled core together with the trial stimulus, but not the raw profile text, so personality-related information at this stage is mediated through the structured core rather than through unrestricted profile conditioning. The decision stage receives only the elicited state(s) and the trial, with no access to either the core or the original profile. Taken together, these stagewise input restrictions create an explicit representational bottleneck at inference time, intended to encourage behavior to depend on intermediate representations rather than on direct access to upstream raw input. A second key choice concerns reuse and compositionality: the personality core is computed once per participant and cached across trials, whereas elicited states are recomputed for each trial or item-level condition as needed. This separation mirrors the CAPS-theoretic distinction between chronic accessible units and their situational activation—a single stable personality foundation gives rise to different internal states under different experimental conditions, which in turn produce different behavioral responses—allowing the pipeline to simultaneously maintain a coherent participant identity across tasks and exhibit appropriate sensitivity when the situational context changes, the two requirements that end-to-end prompting (Eq.~\eqref{eq:e2e}) cannot structurally guarantee.

Finally, the pipeline follows a causal ordering that reflects the temporal and logical structure of human decision-making in experimental settings: who the participant is precedes what the situation activates, which in turn precedes what the participant does. This ordering is not merely a notational convenience. By forcing the model to commit to a personality representation before seeing the task and to commit to an internal state before producing a response, we prevent later stages from retroactively reshaping earlier representations to fit a desired answer. Recent empirical studies on LLM-based social learning have demonstrated that forcing an agent to explicitly articulate its internal cognitive state (e.g., a probabilistic belief) prior to executing a discrete action successfully surfaces latent, human-like cognitive biases that are otherwise smoothed over by end-to-end generation~\citep{arghal2025steering}. Each stage's output is finalized before the next stage begins, making the source of each behavioral decision explicit and traceable.

\section{Experiments}
\subsection{Experimental Setup}\label{sec:alignment-metrics}

\textbf{Study Selection.} We evaluate SPIN on two reconstructed social-psychological study families, with detailed scenario descriptions deferred to Appendix~\ref{app:scenario_info}. The Shafir--Tversky family studies reasoning under uncertainty by holding decision payoffs fixed while varying the informational state~\citep{shafir1992thinking}. It includes Prisoner's Dilemma variants, a computerized Newcomb task, and an information-seeking decision problem, from which we extract discrete observables such as cooperation, one-box choice, and willingness to pay for information. The Prentice--Miller family studies pluralistic ignorance in campus drinking attitudes by separating private attitudes, perceived peer norms, and action-relevant social responses~\citep{prentice1993pluralistic}. Its observables include self--other comfort judgments, perceived norm gaps, dispersion patterns, and downstream behavioral indicators.

\textbf{Data Extraction.} We build benchmark trials from the structured task records of \citet{liu2026humanstudy}\footnote{\url{https://github.com/AISmithLab/HumanStudy-Bench/tree/main}}, aligned with the original articles for the two study families. Each record separates participant-facing protocol fields from reference-outcome fields: the former specify participant setup, condition assignments, trial materials, response options, item order, and measurement scales, whereas the latter store the human findings used only for evaluation. This separation prevents outcome leakage, since prompts expose only participant-facing materials and empirical outcomes remain held out until scoring. The records also define answer-normalization rules that map model outputs to the observables reported in the source studies, including condition-level choice rates for Shafir--Tversky and rating means, gaps, dispersion, and action-related outcomes for Prentice--Miller.

\textbf{Alignment Metrics}. Following recent validation work for LLM-based behavioral simulations~\citep{hullman2026human,cui2025large,park2024generative}, we evaluate alignment at the level of theory-relevant behavioral summaries rather than raw response strings. For each scenario family, model outputs are first normalized into the same observable quantities reported in the human experiments, such as choice rates, and condition gaps. For each scalar metric component $m$, with model summary $a_m$, human target $h_m$, and normalization bound $B_m$, we compute the normalized alignment score as:
\[
  s_m=\operatorname{clip}_{[0,1]}\!\left(1-\frac{|a_m-h_m|}{B_m}\right),
\]
where $s_m$ is the alignment score for component $m$, and $B_m$ is the natural range used for that observable (e.g., $1$ for proportions, $2$ for differences between proportions, $10$ for 11-point rating scales, and $100$ for percentage-scale outcomes). Primary-family scores average their valid components, including the structural ordering checks described in Appendix~\ref{app:alignment_metric}, and study-level alignment is the unweighted mean of primary-family scores. This family-level aggregation prevents high-volume blocks of cells from dominating the headline score. The Shafir--Tversky families are information-seeking calibration (PF-IS), Prisoner's Dilemma triad calibration (PF-PD), and Newcomb calibration (PF-NC); the Prentice--Miller families are mean-level calibration (PF-ML), normative-gap structure (PF-NG), dispersion alignment (PF-DA), and social-action alignment (PF-SA). We provide the concrete components and scales in Appendix~\ref{app:alignment_metric}.

\textbf{Baselines}. We compare SPIN against five prompting-based baselines. \texttt{Blank} provides only the reconstructed experimental task, without any participant-level specification. \texttt{Demographic} augments the task with age, gender, and education attributes sampled from the source study's participant distribution, following prior persona-conditioning and demographic-conditioning work~\citep{argyle2023out,aher2023using,jiang2024personallm}. \texttt{Narrative} replaces this compact profile with a natural-language life background covering personality, routines, and social context, following memory-rich agent simulation~\citep{park2023generative,park2024generative}. We follow the concrete implementation of these three baselines in \citet{liu2026humanstudy}. To control for gains from extra reasoning and staged decomposition, we additionally evaluate \texttt{Demographic+CoT} and \texttt{Narrative+CoT}. These variants use the same participant specifications as their non-CoT counterparts but run in two calls: the first produces an unconstrained natural-language reasoning note about how the participant may interpret the scenario, and the second receives the original input together with this note to produce the required answer. Evaluation uses only the final canonical answer, and these CoT variants introduce no SPIN-specific structure, terminology, or intermediate-state constraints. More details are given in Appendix~\ref{app:baselines}.

\subsection{Main Results}

We evaluate SPIN and baselines on four base LLMs, consisting of two open-source models (DeepSeek~V3.2 and Qwen~3.5-max) and two closed-source models (GPT-5.4-mini and Gemini-3-flash), over the two study families described above. Table~\ref{tab:main_alignment} reports the study-level alignment scores for Shafir--Tversky and Prentice--Miller, along with Avg., their mean over the two headline scores; representative structured outputs are provided in Appendix~\ref{app:parsed_trajectories}.

\begin{table*}[h]
  \centering
  \scriptsize
  \setlength{\tabcolsep}{2.5pt}
  \caption{Behavioral alignment on the experimental scenario families across base-LLMs and prompting variants. Avg. is the mean of the two family headline alignment scores (Align.); primary-family columns are as defined in Section~\ref{sec:alignment-metrics}.  Demo. denotes \texttt{Demographic}, and Narr. denotes \texttt{Narrative}. Within each model block, the highest values in Avg. and Align. columns are shown in \textbf{bold}, and the second-highest values are \underline{underlined}.}
  \label{tab:main_alignment}
  \resizebox{\textwidth}{!}{%
  \begin{tabular}{c|c|c|cccc|ccccc}
    \toprule
    & & & \multicolumn{4}{c|}{Shafir--Tversky (1992)} & \multicolumn{5}{c}{Prentice--Miller (1993)} \\
    \cmidrule(lr){4-7}\cmidrule(lr){8-12}
    Model & Method & Avg.↑ &
    Align. & PF-IS & PF-PD & PF-NC &
    Align. & PF-ML & PF-NG & PF-DA & PF-SA \\
    \midrule
    \multirow{6}{*}{GPT-5.4-mini}
      & SPIN       & \textbf{0.846} & \textbf{0.828} & 0.860 & 0.698 & 0.925 & \textbf{0.863} & 0.955 & 0.857 & 0.854 & 0.784 \\
      & Blank      & 0.704 & 0.574 & 0.302 & 0.694 & 0.725 & 0.834 & 0.911 & 0.798 & 0.865 & 0.760 \\
      & Demo.      & 0.723 & 0.611 & 0.460 & 0.689 & 0.683 & 0.835 & 0.913 & 0.837 & 0.874 & 0.715 \\
      & Demo.+CoT  & 0.793 & 0.735 & 0.677 & 0.703 & 0.825 & 0.852 & 0.925 & 0.885 & 0.859 & 0.739 \\
      & Narr.      & 0.805 & 0.761 & 0.802 & 0.655 & 0.825 & 0.839 & 0.809 & 0.924 & 0.854 & 0.768 \\
      & Narr.+CoT  & \underline{0.811} & \underline{0.773} & 0.850 & 0.675 & 0.795 & \underline{0.849} &0.854 & 0.937 & 0.865 & 0.741 \\
    \midrule
    \multirow{6}{*}{Gemini-3-flash}
      & SPIN       & \textbf{0.875} & \textbf{0.839} & 0.952 & 0.913 & 0.650 & \textbf{0.911} & 0.957 & 0.952 & 0.858 & 0.838 \\
      & Blank      & 0.700 & 0.524 & 0.227 & 0.694 & 0.650 & 0.876 & 0.967 & 0.984 & 0.836 & 0.719 \\
      & Demo.      & 0.793 & 0.701 & 0.810 & 0.642 & 0.650 & \underline{0.885} & 0.941 & 0.949 & 0.843 & 0.807 \\
      & Demo.+CoT  & \underline{0.823} & \underline{0.774} & 0.807 & 0.866 & 0.650 & 0.871 & 0.938 & 0.943 & 0.845 & 0.757 \\
      & Narr.      & 0.714 & 0.557 & 0.386 & 0.635 & 0.650 & 0.870 & 0.861 & 0.872 & 0.863 & 0.884 \\
      & Narr.+CoT  & 0.722 & 0.578 & 0.452 & 0.632 & 0.650 & 0.866 & 0.859 & 0.868 & 0.861 & 0.876 \\
    \midrule
    \multirow{6}{*}{DeepSeek V3.2}
      & SPIN       & \textbf{0.869} & \textbf{0.830} & 0.735 & 0.779 & 0.975 & \textbf{0.908} & 0.976 & 0.974 & 0.900 & 0.782 \\
      & Blank      & 0.714 & 0.582 & 0.282 & 0.813 & 0.650 & 0.847 & 0.911 & 0.791 & 0.894 & 0.792 \\
      & Demo.      & 0.719 & 0.569 & 0.302 & 0.754 & 0.650 & 0.869 & 0.927 & 0.879 & 0.902 & 0.767 \\
      & Demo.+CoT  & \underline{0.770} & \underline{0.659} & 0.352 & 0.824 & 0.800 & \underline{0.881} & 0.955 & 0.919 & 0.926 & 0.724 \\
      & Narr.      & 0.676 & 0.493 & 0.190 & 0.639 & 0.650 & 0.860 & 0.853 & 0.924 & 0.878 & 0.785 \\
      & Narr.+CoT  & 0.706 & 0.559 & 0.340 & 0.663 & 0.675 & 0.852 & 0.855 & 0.922 & 0.890 & 0.742 \\
    \midrule
    \multirow{6}{*}{Qwen 3.5-max}
      & SPIN       & \textbf{0.821} & \textbf{0.830} & 0.823 & 0.768 & 0.900 & \textbf{0.811} & 0.879 & 0.880 & 0.910 & 0.576 \\
      & Blank      & 0.694 & 0.636 & 0.640 & 0.644 & 0.625 & 0.752 & 0.785 & 0.785 & 0.875 & 0.564 \\
      & Demo.      & 0.740 & 0.711 & 0.715 & 0.493 & 0.925 & 0.769 & 0.798 & 0.844 & 0.870 & 0.564 \\
      & Demo.+CoT  & 0.649 & 0.525 & 0.427 & 0.474 & 0.675 & 0.772 & 0.787 & 0.845 & 0.861 & 0.597 \\
      & Narr.      & \underline{0.750} & \underline{0.734} & 0.752 & 0.600 & 0.850 & 0.765 & 0.728 & 0.864 & 0.858 & 0.609 \\
      & Narr.+CoT  & 0.723 & 0.654 & 0.477 & 0.585 & 0.900 & \underline{0.791} & 0.743 & 0.975 & 0.866 & 0.580 \\
    \bottomrule
  \end{tabular}%
  }
\end{table*}

SPIN obtains the highest Avg. score for all four backbones and ranks first on both study-level alignment scores in every model block. A backbone-paired robustness check (Appendix~\ref{app:paired_backbone_test}) shows that SPIN's improvement is positive on every backbone for every baseline in both study families, with 40 out of 40 paired differences positive. The strongest overall result appears with Gemini-3-flash, where SPIN reaches an Avg. of 0.875. The improvement is especially large for DeepSeek~V3.2, where the best baseline reaches 0.770 while SPIN reaches 0.869. Across models, the same qualitative pattern holds: SPIN improves the two-study macro-average without trading off one study family against the other. The primary-family scores indicate where these gains arise. In the Shafir--Tversky family, SPIN most consistently improves the components that require sensitivity to informational condition, especially information seeking and the Prisoner's Dilemma triad. This pattern is consistent with the intended role of SPIN's state-elicitation stage: the model must preserve a stable participant representation while updating its situational interpretation as the opponent information changes. In the Prentice--Miller family, the gains are generally smaller because several baselines already capture much of the aggregate norm structure, but SPIN still obtains the strongest study-level alignment for every backbone. This suggests that the method not only improves the formatting of local responses, but also helps preserve distinctions between private attitudes, perceived norms, and downstream social responses.

These aggregate results support the narrower hypothesis that separating personality compilation, situational state elicitation, and decision readout can improve benchmark-level summary alignment in the socially structured tasks considered here. Notably, SPIN's gains are consistent across all four backbones and both study families, and the largest on condition-sensitive components. The ablation study below tests this hypothesis more directly by removing individual components of SPIN, while the case studies inspect whether the improved scores correspond to interpretable changes in the elicited states rather than to only better final-answer calibration.

\subsection{Ablations}

To isolate which components of SPIN account for its gains, we compare the full pipeline against three targeted ablations under the same backbone, decoding configuration, and answer-normalization protocol (Table~\ref{tab:ablations}). \texttt{SPIN\_wo\_elicitation} removes the situation-specific state-construction stage: the decision module receives the compiled personality core and the trial directly, testing whether an explicit transient state is necessary for condition-sensitive behavior. \texttt{SPIN\_wo\_compilation} removes the task-blind personality compilation stage: instead of first constructing a stable participant core, the elicitation module operates directly on the raw participant profile together with the trial, testing whether explicitly compiling reusable participant structure contributes beyond direct profile conditioning. \texttt{SPIN\_freeform\_intermediate} preserves a multi-stage procedure but replaces SPIN's fixed trait/state schema with an unconstrained natural-language intermediate analysis, testing whether the observed gains arise from structured intermediate representations rather than merely from inserting an additional reasoning step.

\begin{table*}[h]
  \centering
  \scriptsize
  \setlength{\tabcolsep}{3.5pt}
  \caption{Ablation comparison across base models. Avg. = mean of the two family-level headline scores; S--T = Shafir--Tversky (1992); P--M = Prentice--Miller (1993). Row names: SPIN = full method; w/o~elicitation = \texttt{SPIN\_wo\_elicitation}; w/o~compilation = \texttt{SPIN\_wo\_compilation}; freeform~interm. = \texttt{SPIN\_freeform\_intermediate}. Within each model block, the highest values are shown in \textbf{bold}, and the second-highest values are \underline{underlined}.}
  \label{tab:ablations}
  \resizebox{\textwidth}{!}{%
  \begin{tabular}{l|ccc|ccc|ccc|ccc}
    \toprule
    & \multicolumn{3}{c|}{GPT-5.4-mini} &
      \multicolumn{3}{c|}{Gemini-3-flash} &
      \multicolumn{3}{c|}{DeepSeek V3.2} &
      \multicolumn{3}{c}{Qwen 3.5-max} \\
    \cmidrule(lr){2-4}\cmidrule(lr){5-7}\cmidrule(lr){8-10}\cmidrule(lr){11-13}
    Method
    & Avg.↑ & S--T & P--M
    & Avg.↑ & S--T & P--M
    & Avg.↑ & S--T & P--M
    & Avg.↑ & S--T & P--M \\
    \midrule
    SPIN
    & \textbf{0.845} & \textbf{0.828} & \textbf{0.863}
    & \textbf{0.875} & \textbf{0.839} & \textbf{0.911}
    & \textbf{0.869} & \textbf{0.830} & \textbf{0.908}
    & \textbf{0.821} & \textbf{0.830} & \underline{0.811} \\
    w/o~elicitation
    & 0.809 & \underline{0.790} & 0.828
    & 0.820 & 0.764 & 0.876
    & 0.818 & 0.773 & 0.862
    & 0.776 & 0.757 & 0.794 \\
    w/o~compilation
    & \underline{0.812} & 0.767 & \underline{0.856}
    & \underline{0.852} & \underline{0.805} & \underline{0.898}
    & \underline{0.848} & \underline{0.795} & \underline{0.901}
    & \underline{0.798} & \underline{0.813} & 0.782 \\
    freeform~interm.
    & 0.787 & 0.739 & 0.834
    & 0.816 & 0.742 & 0.890
    & 0.801 & 0.724 & 0.878
    & 0.747 & 0.678 & \textbf{0.815} \\
    \bottomrule
  \end{tabular}%
  }
\end{table*}

The two removal ablations show that both components are useful, with elicitation playing the larger role. Removing either the stable participant core or the situation-specific elicitation stage lowers Avg. for all four backbones, indicating that neither component is redundant. The drop is consistently larger for \texttt{SPIN\_wo\_elicitation} than for \texttt{SPIN\_wo\_compilation} (0.036--0.055 vs. 0.021--0.033), especially on Shafir--Tversky. This supports the intended role of elicitation: converting a stable participant representation into a transient state that tracks condition-specific information. The smaller but systematic degradation from \texttt{SPIN\_wo\_compilation} suggests that direct raw-profile conditioning is not a full substitute for an explicit task-blind participant core.

The \texttt{SPIN\_freeform\_intermediate} ablation distinguishes SPIN from generic multi-step prompting. Despite preserving the staged procedure, replacing the fixed trait/state schema with unconstrained natural-language reasoning reduces Avg. for every backbone, with the largest drops on Shafir--Tversky (0.089--0.152). This suggests that SPIN's gains depend on a structured intermediate state, not merely an extra reasoning pass. Although \texttt{SPIN\_freeform\_intermediate} remains competitive on Prentice--Miller and slightly exceeds SPIN for Qwen~3.5-max, this exception is not robust, indicating that structured intermediates matter most when alignment requires both stable identity and condition-sensitive updating.

\subsection{Behavioral Calibration by State Elicitation}

\begin{wrapfigure}{r}{0.50\textwidth}
  \centering
  \vspace{-1.8em}
  \includegraphics[width=\linewidth]{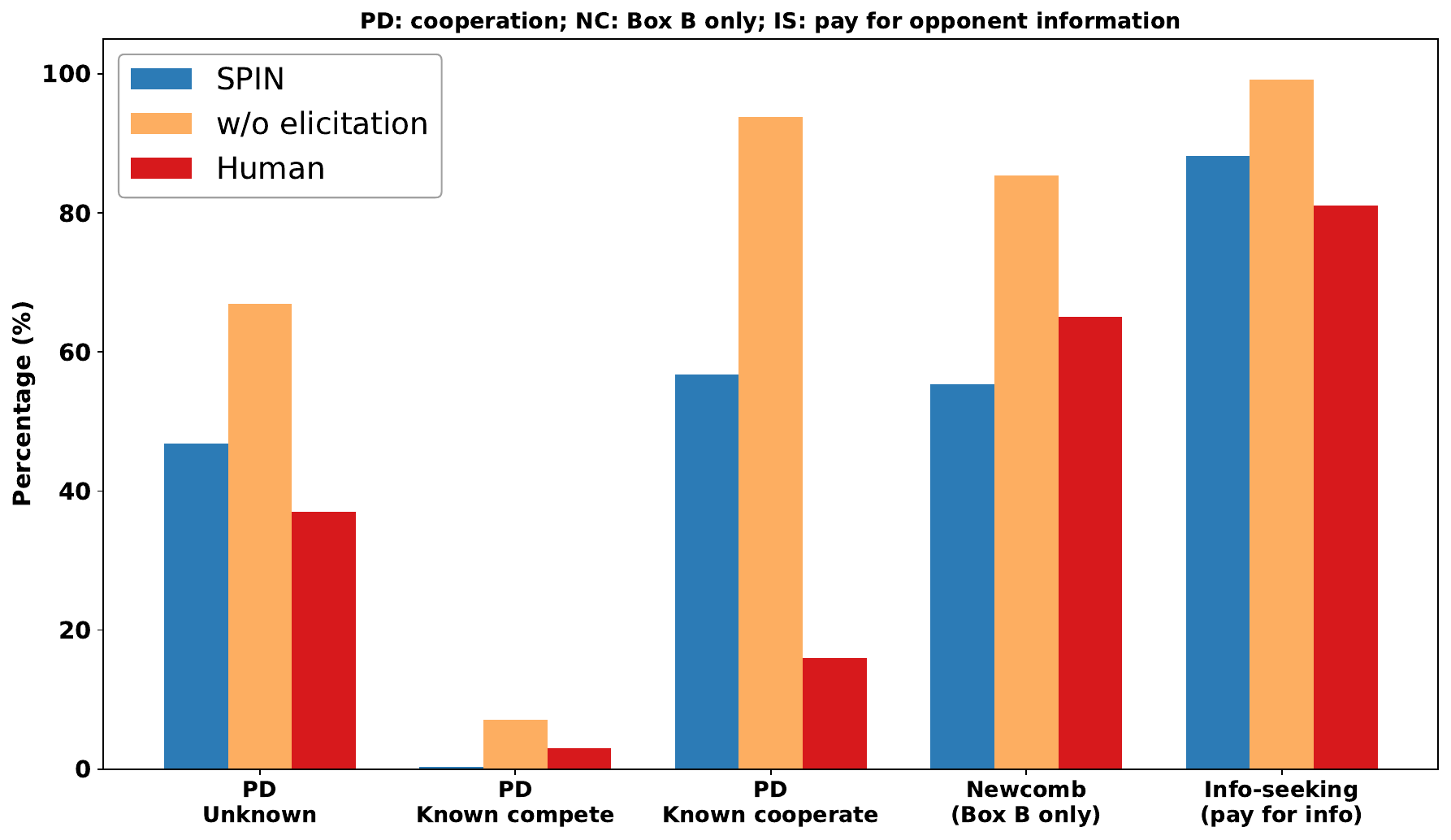}
  \caption{Behavioral consequences of removing state elicitation. Each group reports the percentage of the task-specific focal action: cooperation in Prisoner's Dilemma, choosing Box~B only in Newcomb's problem, and paying for opponent information in the information-seeking condition.}
  \label{fig:st_bar}
  \vspace{-1.2em}
\end{wrapfigure}

We next examine whether the elicited state layer has functional consequences for downstream behavior. Figure~\ref{fig:st_bar} compares the full SPIN pipeline with the \texttt{w/o~elicitation} ablation across five condition groups, using the task-specific focal action in each: cooperation in the Prisoner's Dilemma, choosing Box~B only in Newcomb's problem, and paying for opponent information in the information-seeking condition. Removing state elicitation consistently shifts the model toward higher focal-action rates---especially in the known-cooperate condition, Newcomb's problem, and information seeking, where the ablated variant approaches a near-uniform high-response policy---while the full pipeline produces more moderated behavior generally closer to the human target. These patterns suggest that state elicitation does not uniformly increase human-like responding, but rather calibrates the final readout to the inferred situational state, acting as a behavioral regularizer that ties choices more closely to condition-specific appraisals; Appendix~\ref{app:case_study_1} provides additional details and an interpretation.

\subsection{Aggregate State Sensitivity}

\begin{wrapfigure}{r}{0.50\textwidth}
  \centering
  \vspace{-1.8em}
  \includegraphics[width=\linewidth]{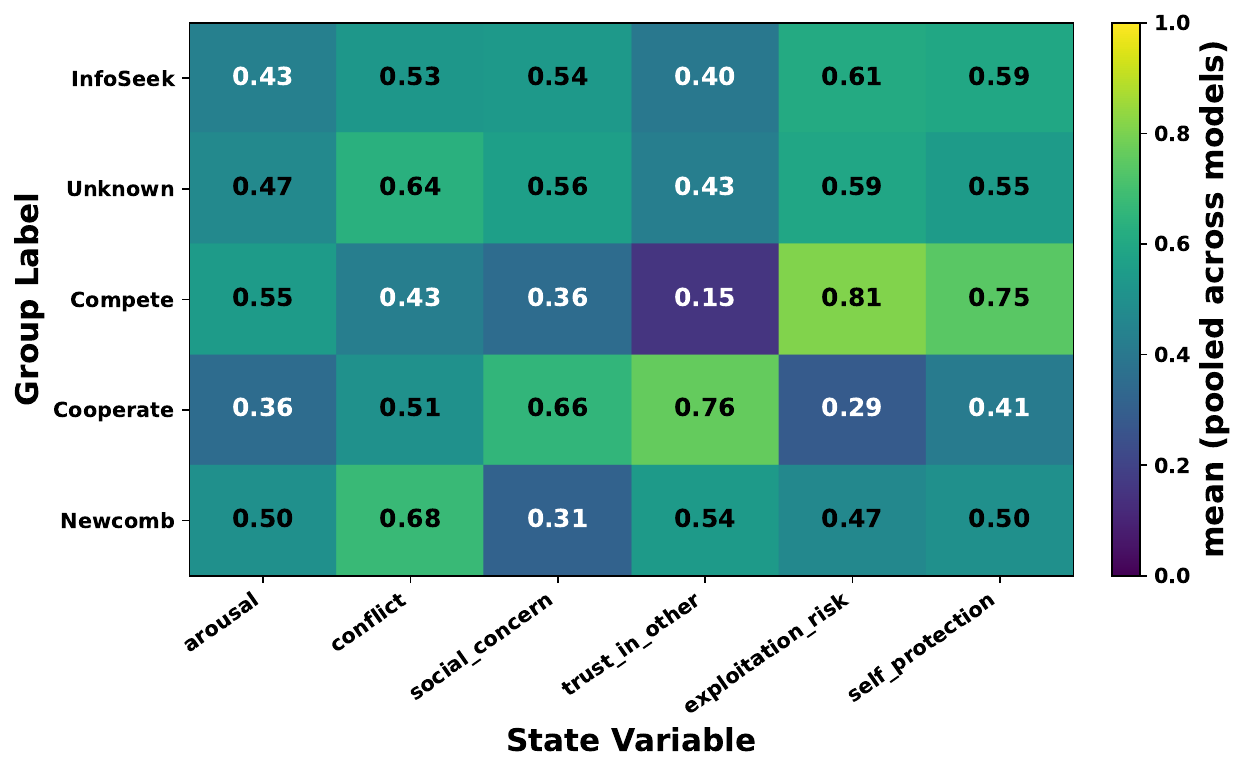}
  \caption{Aggregate state sensitivity in the Shafir--Tversky family. The heatmap reports mean elicited state values across five condition groups, pooled over participants and four base-LLMs.}
  \label{fig:st_heatmap}
  \vspace{-1em}
\end{wrapfigure}

We analyze whether SPIN's structured intermediate states respond systematically to condition changes in the Shafir–Tversky family. Figure~\ref{fig:st_heatmap} reports the mean values obtained from the six state variables across five condition groups, pooled over participants and backbones. The resulting patterns are consistent with the experimental manipulations: within the Prisoner's Dilemma triad, the compete condition yields the lowest trust and the highest exploitation risk and self-protection, the cooperate condition shows the opposite profile, and the unknown condition lies between these extremes with elevated conflict reflecting opponent ambiguity. The information-seeking and Newcomb conditions also induce distinct state profiles rather than collapsing into a shared generic response. Overall, when the informational structure of the task changes, the elicited states shift in interpretable directions rather than merely restating participant profiles. The analogous pattern in the Prentice–Miller family (Figure~\ref{fig:pm_heatmap}, Appendix~\ref{app:case_study_2}) shows systematic differentiation between conditions centered on private attitudes, perceived peer norms, and downstream social alignment.

\section{Conclusion}
This paper formalized social judgment and decision alignment as matching simulated behavioral summaries to theory-relevant human targets, and introduced SPIN, a zero-shot CAPS-inspired pipeline that stages task-blind personality compilation, condition-specific state elicitation, and state-grounded readout. Across two reconstructed study families and four base LLMs, SPIN improves alignment over blank, demographic, narrative, and their chain-of-thought augmented variants. Ablations show that task-blind personality compilation and structured state elicitation both contribute, while state analyses suggest that intermediate states shift interpretably across informational and normative conditions rather than as generic reasoning traces. SPIN should therefore be viewed as a structured inference method for behavioral alignment, not as a validated psychological measurement model; broader study coverage and cost-matched baselines remain important directions.

% \begin{ack}
% Use unnumbered first level headings for the acknowledgments. All acknowledgments

% Do {\bf not} include this section in the anonymized submission, only in the final paper. You can use the \texttt{ack} environment provided in the style file to automatically hide this section in the anonymized submission.
% \end{ack}

\bibliographystyle{unsrtnat}
\bibliography{references}

%%%%%%%%%%%%%%%%%%%%%%%%%%%%%%%%%%%%%%%%%%%%%%%%%%%%%%%%%%%%
\newpage
\appendix

\section{Related Work}
\label{app:related}

\subsection{LLM Social Simulation: Capabilities, Limitations, and Evaluation}
\label{app:rw-simulation}

Large language models have increasingly been used as synthetic respondents in social and behavioral research. Early work showed that LLMs can approximate human response patterns in economic games~\citep{horton2023large}, replicate directional effects in classic psychological experiments~\citep{aher2023using}, and generate subgroup-sensitive survey responses when conditioned on demographic or partisan cues~\citep{argyle2023out}, motivating broader interest in LLMs as scalable surrogates for human participants~\citep{dillion2023can,demszky2023using}. Subsequent work extended this evidence in both scale and ecological richness. Large-scale replication studies report strong correspondence between simulated and human treatment effects across preregistered or scenario-based experiments~\citep{hewitt2024predicting,cui2025large}; agent-based approaches enrich the simulation with persistent memory, interviews, social interaction, or individualized histories~\citep{park2023generative,park2024generative}; and specialized or hybrid systems further combine LLMs with human-experiment corpora or causal models~\citep{binz2025foundation,manning2024automated}. These results establish that LLMs can produce surprisingly plausible aggregate behavioral patterns across diverse social and cognitive settings.

At the same time, a parallel literature cautions against treating such correspondence as sufficient evidence of human-like behavior. From a causal-inference perspective, apparent LLM--human agreement may reflect prompt artifacts rather than genuine behavioral alignment~\citep{gui2023challenge}. Empirical validation studies show that LLM-generated samples can compress population variance, miscalibrate effect magnitudes, distort causal estimates, and fail to recover subgroup heterogeneity~\citep{bisbee2024synthetic,gao2025take}. More recent frameworks emphasize that reproducing aggregate means is insufficient without also matching distributional structure, cross-condition contrasts, and the intended evidential role of the original human study~\citep{hullman2026human}. Individual-level simulation raises further concerns: personas can drift from assigned characteristics over interaction~\citep{abdulhai2025consistently}, and broader position papers identify diversity, bias, sycophancy, alienness, and generalization as unresolved challenges for LLM social simulation~\citep{anthis2025llm}.

These limitations motivate more structured evaluation targets. Early studies often emphasized aggregate correlations or effect-direction recovery~\citep{horton2023large,aher2023using}, whereas later work distinguishes surface-level response similarity from stronger claims about cognitive or process equivalence~\citep{dillion2023can}. Benchmark efforts such as HumanStudy-Bench separate participant-facing protocols from held-out reference outcomes and normalize model outputs into the observables reported in source studies~\citep{liu2026humanstudy}; related large-scale replications also score theory-relevant behavioral summaries, including choice rates, rating statistics, and treatment contrasts~\citep{cui2025large}. SPIN follows this evaluation view: the target is not to recover full individual psychologies, but to improve benchmark-level alignment with the summaries that social and behavioral studies are designed to measure.

\subsection{Persona Conditioning and Multi-Step Prompting}
\label{app:rw-persona}

A common way to steer LLM behavior toward a target population or individual is persona conditioning: augmenting the prompt with demographic attributes, personality descriptors, social identities, or biographical context. Such conditioning can yield subgroup-level opinion distributions that track survey marginals~\citep{argyle2023out}, induce trait-consistent text under explicit personality descriptors~\citep{jiang2024personallm}, and produce reliable inventory-style responses under generic persona prompts~\citep{serapio2023personality}. These results suggest that LLMs encode enough social and psychometric regularities to generate persona-consistent surface behavior. However, persona prompting remains brittle for social simulation. Demographic steering reduces, but does not eliminate, misalignment between LM-generated and human opinion distributions~\citep{santurkar2023whose}; in-context impersonation can recover some human-like patterns while amplifying sociocultural biases~\citep{salewski2023context}; and richer character modeling often requires fine-tuning or explicit consistency optimization~\citep{shao2023character,abdulhai2025consistently}. More broadly, persona-prompted models can compress inter-individual variability and generate stereotyped responses, especially for underrepresented groups~\citep{dominguez2024questioning}. Thus, simply adding more profile content does not guarantee stable or condition-sensitive behavioral alignment.

A separate prompting literature improves reasoning by decomposing generation into intermediate steps. Chain-of-thought (CoT) prompting and its zero-shot variant elicit intermediate rationales before final answers~\citep{wei2022chain,kojima2022large}; self-consistency, least-to-most prompting, decomposed prompting, and tree-of-thought search further improve reliability by sampling, decomposing, routing, or searching over reasoning paths~\citep{wang2022self,zhou2022least,khot2022decomposed,yao2023tree}. In LLM social simulation, \citet{kong2026improving} similarly distinguish context formation from context navigation, showing that complex decision environments often require both accurate task representation and guided reasoning. These methods motivate controls for whether SPIN's gains come merely from additional reasoning. Our comparison with CoT baselines and free-form intermediate ablations addresses this possibility: generic multi-step reasoning can add computation, but it does not enforce a reusable participant representation or a condition-specific state bottleneck. This distinction is consistent with recent evidence that explicitly articulating internal cognitive states before action can reveal human-like biases otherwise smoothed over by end-to-end generation~\citep{arghal2025steering}.

\subsection{Psychological Theory as Computational Scaffolding for LLM Behavior}
\label{app:rw-psych}

Psychological theory has been used to study LLMs along two related directions. One line treats LLMs as objects of psychological analysis: models are administered cognitive tasks or personality probes to assess whether their outputs exhibit human-like biases, heuristics, traits, or learning patterns~\citep{binz2023using,hagendorff2023machine,hagendorff2023human}. Benchmarks such as CogBench systematize this approach by comparing LLM behavior against human baselines across cognitive experiments~\citep{coda2024cogbench}. Cognitive models have also been used to interpret alignment-relevant value trade-offs in LLMs, such as how models balance informational, social, and presentational utilities in pragmatic communication~\citep{murthycognitive}. These studies show that psychological paradigms are useful diagnostic tools, but they primarily evaluate the behavior of the existing model rather than specifying how behavioral responses should be generated.

A second line uses psychological concepts as design constraints for language agents. CoALA proposes a broad cognitive-architecture view of language agents, organizing memory, action, and decision modules through concepts from cognitive science and symbolic AI~\citep{sumers2023cognitive}. Other work analyzes LLM decisions through dual-process theory~\citep{brady2025dual} or manipulates affective framing, showing that anxiety-like prompts can change exploratory behavior and induce systematic biases~\citep{coda2023inducing}. Recent cognitively faithful decision models further argue that alignment should capture not only choices but also the heuristic or rule-like processes by which humans transform information before deciding~\citep{cousins2025towards}. More directly related to social simulation, explicit belief elicitation before action can surface latent human-like biases that are otherwise smoothed over by end-to-end generation~\citep{arghal2025steering}, while individualized backstory generation can improve value-grounded responses~\citep{du2025simvbg}. Related work on modeling others' minds as code represents routine behavior as executable scripts inferred from sparse observations, suggesting another route for imposing structured, interpretable behavioral representations~\citep{jha2025modeling}. Together, these findings suggest that intermediate psychological or computational representations can affect downstream behavior rather than serving only as post-hoc explanations.

Personality theory provides a further motivation to separate a stable individual structure from situation-specific activation. The Cognitive-Affective Personality System (CAPS) explains personality in terms of stable if--then patterns: cognitive-affective units such as encodings, expectancies, goals, values, and self-regulatory plans are selectively activated by situational features~\citep{caps-1995,mischel2008toward}. This account is especially relevant to participant simulation, where a model must preserve a coherent participant identity while producing differentiated responses to changes in informational state, social target, or normative framing. However, most LLM simulation approaches still rely on either enriched persona descriptions or generic reasoning traces, leaving open the question of how to operationalize personality theory as a structured inference process for benchmark-level behavioral alignment.

\section{Trait and State Dimension Rationale}
\label{app:dimension_justification}

The SPIN intermediate vocabulary is part of the method specification. It is designed as a compact interface for social judgment and strategic-interaction settings rather than as a direct operationalization of a single psychometric scale, following the CAPS-inspired distinction between chronic dispositions and situation-specific activation used in the main method section.

The five trait dimensions span the primary stable behavioral axes implicated in cooperation, fairness, and strategic reasoning: sensitivity to unequal outcomes, responsiveness to social expectations, tendency to reciprocate or punish, preference for resolving ambiguity, and baseline inclination to avoid exploitation or loss. The six state dimensions capture transient psychological activations that a specific scenario may trigger: emotional intensity, internal motivational tension, activated other-regarding concern, momentary trust, situation-specific vulnerability appraisal, and defensive impulse. Together, the vocabulary is designed to cover the main sources of behavioral variation present in the social-dilemma and judgment-under-uncertainty tasks included in our evaluation. Alternative vocabularies could be substituted within the same pipeline when extending to other task families; evaluating the sensitivity to such vocabulary choices is left to future work. Table~\ref{tab:dimension_rationale} lists each dimension together with the established psychological construct that motivates its inclusion. SPIN dimensions are not claimed to be equivalent to any single scale, but are designed to cover the corresponding behavioral axis.

\begin{table}[h]
  \centering
  \small
  \setlength{\tabcolsep}{27pt}
  \caption{SPIN trait and state dimensions with motivating constructs.}
  \label{tab:dimension_rationale}
  \begin{tabular}{lll}
    \toprule
    Dimension & Role & Related construct \\
    \midrule
    fairness sensitivity       & Trait & Inequity aversion \citep{fehr1999theory} \\
    norm sensitivity           & Trait & Social norm compliance \citep{bicchieri2005grammar} \\
    reciprocity orientation    & Trait & Social Value Orientation \citep{murphy2011measuring, trivers1971evolution} \\
    uncertainty aversion       & Trait & Ambiguity aversion \citep{ellsberg1961risk} \\
    self-protection tendency   & Trait & Loss aversion \citep{kahneman2013prospect} \\
    \midrule
    arousal                    & State & Appraisal theory \citep{lazarus1991emotion} \\
    conflict                   & State & Decisional conflict \citep{janis1977decision} \\
    social concern             & State & Other-regarding preferences \citep{fehr1999theory} \\
    trust in other             & State & Trust game \citep{berg1995trust} \\
    perceived exploitation risk & State & Betrayal aversion \citep{bohnet2004trust} \\
    self-protection drive      & State & Threat appraisal \citep{lazarus1991emotion} \\
    \bottomrule
  \end{tabular}
\end{table}

\section{Experimental Scenario Information}
\label{app:scenario_info}

This appendix section summarizes the experimental scenarios considered in the main evaluation.

\subsection{Shafir and Tversky (1992)}

One experimental source used in this paper is Shafir and Tversky's \emph{Thinking through Uncertainty: Nonconsequential Reasoning and Choice} \citep{shafir1992thinking}. This article is a canonical study of the disjunction effect and related violations of Savage's sure-thing principle. It is directly relevant to our setting because it provides behaviorally matched paradigms in which the underlying incentive structure is held fixed while the informational state changes, making it possible to test whether a model preserves stable decision tendencies while still producing condition-sensitive variation under uncertainty.

Our benchmark instantiates three tasks from this scenario family. The first is a one-shot Prisoner's Dilemma presented in three matched informational conditions: the participant does not know the other player's action, is told that the other player cooperated, or is told that the other player competed. This task is central because it isolates the effect of uncertainty while holding payoffs fixed; the human target includes the characteristic reversal in which cooperation is highest in the unknown condition. The second task is a computerized version of Newcomb's Problem, which probes whether the agent prefers the diagnostically attractive one-box option over the dominance-consistent two-box option. The third is an information-seeking Prisoner's Dilemma variant in which the participant may pay to reveal the opponent's action before deciding, making uncertainty itself an explicit object of choice.

Together, these tasks define the Shafir--Tversky scenario family in our benchmark as a compact evaluation setting centered on reasoning under uncertainty. The family is useful not only because it provides measurable aggregate targets, but because it probes whether a model captures the qualitative reorganization of preference induced by informational uncertainty across closely related decision formats.

\subsection{Prentice and Miller (1993)}

Another experimental source used in this paper is Prentice and Miller's \emph{Pluralistic Ignorance and Alcohol Use on Campus: Some Consequences of Misperceiving the Social Norm} \citep{prentice1993pluralistic}. This article studies how individuals misperceive the prevailing social norm and how those misperceptions shape private attitudes, behavioral adjustment, and social alienation. Its central claim is that students often believe that they are more personally uncomfortable with campus drinking culture than their peers, even when many peers privately hold similar reservations. The article is therefore directly relevant to our setting because it separates private attitude, perceived norm, and downstream behavior, which closely matches the kind of latent mediation structure that our framework aims to model.

This article is especially useful for social-psychological behavior alignment because it does not stop at documenting a static self--other gap. It connects misperception to inferences about both abstract campus-level norms and the attitudes of proximate others, to beliefs about how homogeneous peers appear relative to the self, and to behavioral sequelae when people feel deviant from what they believe others believe. Our benchmark instantiates a cross-sectional protocol from this scenario family built around the following substantive ingredients. Participants report private comfort with campus drinking alongside an estimate of the typical peer's comfort. The materials support evaluation of pluralistic ignorance at the level of means and of dispersion: the tendency for attributed peer attitudes to be both shifted and less variable than self-ratings. The protocol also elicits comfort judgments about the self, friends, and the average student under counterbalanced question order, so that apparent gaps are not artifacts of a fixed item sequence. This probes whether a model represents local and global social targets coherently when both are available. Finally, the materials include a policy-relevant block modeled on campus restrictions on alcohol service. After eliciting how participants read their standing relative to others' attitudes, the task measures willingness to undertake collective action and related outcomes associated with social alienation, connecting perceived normative discrepancy to inhibition of public engagement rather than to attitude reports alone.

Together, these ingredients define the Prentice--Miller scenario family as operationalized in our benchmark. They foreground the distinction between private attitudes and attributed norms, the structure of self--other gaps across social targets, second-order beliefs about the variability of others, and behavioral correlates of feeling misaligned with the perceived crowd. That makes the family well suited for testing whether a model represents not only what an agent personally believes but also what it attributes to others and how those attributions reorganize expressed behavior.

\section{Metric Design}
\label{app:alignment_metric}

This appendix gives the concrete form of the alignment metrics used for the two scenario families discussed above, grounded in the experimental findings reported by Shafir and Tversky \citep{shafir1992thinking} and Prentice and Miller \citep{prentice1993pluralistic}. For both studies, the evaluator maps model outputs to benchmark-level behavioral variables and compares the resulting empirical summaries with human targets from the source experiments.
Let $z_m$ denote a scalar behavioral summary for the metric component $m$: a choice proportion, a rating mean, a standard deviation, or a contrast between two such quantities. Let $a_m$ be the model estimate, $h_m$ the human target, and $B_m$ be the bound used to normalize the error. The component score is
\[
  S(m)=\operatorname{clip}_{[0,1]}\!\left(1-\frac{|a_m-h_m|}{B_m}\right).
\]
The bound $B_m$ is chosen from the measurement scale of the compared observable: $B_m=1$ for proportions, $B_m=2$ for differences between proportions, $B_m=10$ for 11-point rating means, rating standard deviations, and rating gaps, and $B_m=100$ for percentage-scale outcomes. For pairwise ordering components over a set of cells $\mathcal{I}$, we compute
\[
  O(\mathcal{I})=
  \frac{1}{|\mathcal{P}|}\sum_{(i,j)\in\mathcal{P}}
  \mathbf{1}\!\left[\operatorname{sign}(a_i-a_j)=\operatorname{sign}(h_i-h_j)\right],
  \qquad
  \mathcal{P}=\{(i,j): i,j\in\mathcal{I}, i<j\}.
\]
Each primary family score is the arithmetic mean of its valid component scores, and the study-level headline score is the unweighted mean of the primary family scores. Components that cannot be computed because no valid model response is available are omitted from that mean; auxiliary collapse diagnostics are reported separately and do not enter the headline alignment score.

\subsection{Shafir and Tversky (1992)}

For the uncertainty-reasoning scenario family, the evaluator canonicalizes each response into a benchmark-facing decision label, such as \texttt{cooperate}, \texttt{compete}, \texttt{box\_b\_only}, \texttt{both\_boxes}, \texttt{pay}, or \texttt{no\_pay}, and then aggregates these labels into empirical choice rates. The Shafir--Tversky headline score is
\[
  A_{\mathrm{ST}}=\frac{1}{3}\left(F_{\mathrm{IS}}+F_{\mathrm{PD}}+F_{\mathrm{NC}}\right).
\]

\paragraph{PF-IS: information-seeking calibration.}
Let $a_{\mathrm{pay}}$ be the proportion of the model that chooses to pay for information, computed among the valid \texttt{pay}/\texttt{no\_pay} responses, and let $h_{\mathrm{pay}}$ be the corresponding human proportion. The information-seeking family contains one component:
\[
  F_{\mathrm{IS}}=S(\mathrm{pay}), \qquad B_{\mathrm{pay}}=1.
\]

\paragraph{PF-NC: Newcomb calibration.}
Let $a_{\mathrm{B}}$ be the model proportion selecting \texttt{box\_b\_only} among valid \texttt{box\_b\_only}/\texttt{both\_boxes} responses, and let $h_{\mathrm{B}}$ be the human rate of one-box. The Newcomb family is
\[
  F_{\mathrm{NC}}=S(\mathrm{box\_b\_only}), \qquad B_{\mathrm{box\_b\_only}}=1.
\]

\paragraph{PF-PD: Prisoner's Dilemma triad calibration.}
Let $r_U$, $r_C$, and $r_K$ denote the cooperation rates in the unknown, known-cooperate, and known-compete conditions, respectively, with model and human versions $a_\cdot$ and $h_\cdot$. The family includes the three level components $S(r_U)$, $S(r_C)$, and $S(r_K)$ with $B=1$, two condition-gap components,
\[
  S(r_U-r_K), \qquad S(r_U-r_C),
\]
with $B=2$, and one ordering component $O(\{U,C,K\})$. The family score is the mean of these six components:
\[
  F_{\mathrm{PD}}=\operatorname{mean}\{S(r_U),S(r_C),S(r_K),S(r_U-r_K),S(r_U-r_C),O(\{U,C,K\})\}.
\]
This separates marginal cooperation calibration from the structural pattern emphasized by the original study.

\subsection{Prentice and Miller (1993)}

For the norm-misperception scenario family, the evaluator extracts scalar rating responses and compares them with the human summary statistics reported in the benchmark. The Prentice--Miller headline score is
\[
  A_{\mathrm{PM}}=\frac{1}{4}\left(F_{\mathrm{ML}}+F_{\mathrm{NG}}+F_{\mathrm{DA}}+F_{\mathrm{SA}}\right).
\]

\paragraph{PF-ML: mean-level calibration.}
This family compares directly observed rating means. It includes Study~1 self-comfort and average-student estimates, plus Study~2 means for self, friend, and average-student targets under both self-first and average-first question orders. If $\mathcal{M}$ is this set of mean components, then
\[
  F_{\mathrm{ML}}=\operatorname{mean}_{m\in\mathcal{M}} S(m), \qquad B_m=10.
\]

\paragraph{PF-NG: normative-gap structure.}
This family compares self--other contrasts and target orderings. It includes the Study~1 average-minus-self gap, the Study~2 friend-minus-self and average-minus-friend gaps under each order condition, and the pairwise ordering agreement among self, friend, and average-student means within each order condition:
\[
  \begin{aligned}
  F_{\mathrm{NG}}=\operatorname{mean}\{&
  S(g_1), S(g_{2,o}^{\mathrm{friend-self}}),
  S(g_{2,o}^{\mathrm{average-friend}}),\\
  &O_o(\{\mathrm{self},\mathrm{friend},\mathrm{average}\})
  : o\in\{\mathrm{self\mbox{-}first},\mathrm{average\mbox{-}first}\}\},
  \end{aligned}
\]
where $B=10$ for the rating-gap components. These terms operationalize pluralistic ignorance as both gap magnitude and the ordering of perceived social targets.

\paragraph{PF-DA: dispersion alignment.}
This family uses Study~1 standard deviations for self-ratings and average-student estimates, together with their SD gap:
\[
  F_{\mathrm{DA}}=\operatorname{mean}\{S(\sigma_{\mathrm{self}}),S(\sigma_{\mathrm{average}}),
  S(\sigma_{\mathrm{self}}-\sigma_{\mathrm{average}})\}, \qquad B=10.
\]
It captures the variance pattern behind the ``illusion of universality,'' namely that attributed peer attitudes are less variable than private self-ratings.

\paragraph{PF-SA: social-action alignment.}
This family compares the downstream policy-related outcomes in the Study~4. For each perceived-deviance group $g\in\{\mathrm{others\ more\ negative},\mathrm{others\ the\ same}\}$ and outcome $y\in\{\mathrm{signatures},\mathrm{hours},\mathrm{reunions}\}$, the evaluator scores the group mean $S(\mu_{g,y})$. It also scores the group contrast $S(\mu_{\mathrm{same},y}-\mu_{\mathrm{more\ negative},y})$ for each outcome:
\[
  F_{\mathrm{SA}}=\operatorname{mean}_{y}
  \{S(\mu_{\mathrm{more\ negative},y}),S(\mu_{\mathrm{same},y}),
  S(\mu_{\mathrm{same},y}-\mu_{\mathrm{more\ negative},y})\}.
\]
The bound is $B=100$ for the signatures and the reunion percentages and $B=10$ for hours. The benchmark scores only instruments that can be aligned directly to published marginal summary statistics; experimental sections without such comparables are left out of the scored protocol.

\section{Backbone-Paired Robustness Test}
\label{app:paired_backbone_test}

As an auxiliary robustness check, we test whether SPIN's advantage over each prompting baseline is consistent across the four base-LLMs used in the main evaluation. The unit of analysis is the backbone, not the individual simulated trial. For each study family and each baseline, we form paired differences
\[
  d_b = A_{\mathrm{SPIN},b} - A_{\mathrm{baseline},b},
\]
where $A_{\cdot,b}$ is the study-level alignment score for backbone $b$. This gives four paired differences per comparison, corresponding to GPT-5.4-mini, Gemini-3-flash, DeepSeek~V3.2, and Qwen~3.5-max. We report the mean paired difference, a two-sided 95\% confidence interval for that mean, and a one-sided paired $t$-test for the directional hypothesis that SPIN improves alignment over the baseline. The test is equivalent to a one-sample $t$-test on the four differences $d_b$.

This analysis should be interpreted conservatively. With only four paired units, the $t$-test has low power and its normality approximation is fragile; conversely, the exact one-sided sign test has a minimum possible $p$-value of $1/16=0.0625$ when all four differences are positive. We therefore use this check only as a model-level robustness diagnostic, not as the primary evidence for behavioral alignment. Table~\ref{tab:paired_backbone_robustness} shows that SPIN's advantage is positive for every baseline on every backbone in both study families.

\begin{table*}[h]
  \centering
  \scriptsize
  \setlength{\tabcolsep}{3pt}
  \caption{Backbone-paired robustness check. Each $\Delta$ is the paired difference SPIN minus baseline in study-level alignment score for a single backbone. Mean $\Delta$, 95\% CI, and one-sided paired $t$-test use the four backbone-level paired differences. The exact one-sided sign-test $p$-value is 0.0625 for every row because all four differences are positive and $n=4$.}
  \label{tab:paired_backbone_robustness}
  \resizebox{\textwidth}{!}{%
  \begin{tabular}{llrrrrrrrr}
    \toprule
    Study family & Baseline
    & $\Delta_{\mathrm{GPT}}$ & $\Delta_{\mathrm{Gemini}}$
    & $\Delta_{\mathrm{DeepSeek}}$ & $\Delta_{\mathrm{Qwen}}$
    & Mean $\Delta$ & 95\% CI & $t(3)$ & $p_{\mathrm{one}}$ \\
    \midrule
    \multirow{5}{*}{Shafir--Tversky}
      & Blank     & 0.254 & 0.315 & 0.248 & 0.194 & 0.253 & [0.174, 0.332] & 10.21 & 0.001 \\
      & Demo.     & 0.217 & 0.138 & 0.261 & 0.119 & 0.184 & [0.078, 0.290] & 5.51 & 0.006 \\
      & Demo.+CoT & 0.093 & 0.065 & 0.171 & 0.305 & 0.158 & [-0.013, 0.330] & 2.95 & 0.030 \\
      & Narr.     & 0.067 & 0.282 & 0.337 & 0.096 & 0.195 & [-0.018, 0.409] & 2.92 & 0.031 \\
      & Narr.+CoT & 0.055 & 0.261 & 0.271 & 0.176 & 0.191 & [0.032, 0.350] & 3.81 & 0.016 \\
    \midrule
    \multirow{5}{*}{Prentice--Miller}
      & Blank     & 0.029 & 0.035 & 0.061 & 0.059 & 0.046 & [0.020, 0.072] & 5.62 & 0.006 \\
      & Demo.     & 0.028 & 0.026 & 0.039 & 0.042 & 0.034 & [0.021, 0.046] & 8.51 & 0.002 \\
      & Demo.+CoT & 0.011 & 0.040 & 0.027 & 0.039 & 0.029 & [0.008, 0.051] & 4.33 & 0.011 \\
      & Narr.     & 0.024 & 0.041 & 0.048 & 0.046 & 0.040 & [0.022, 0.057] & 7.29 & 0.003 \\
      & Narr.+CoT & 0.014 & 0.045 & 0.056 & 0.020 & 0.034 & [0.002, 0.066] & 3.37 & 0.022 \\
    \bottomrule
  \end{tabular}%
  }
\end{table*}

\subsection{Behavioral Consequences of Removing State Elicitation}
\label{app:case_study_1}

Figure~\ref{fig:st_bar} compares the full SPIN pipeline, the \texttt{w/o~elicitation} ablation, and the original human targets across the five Shafir--Tversky condition groups. The plotted quantity is always a condition-level response rate in $[0,1]$: cooperation rate for the three Prisoner's Dilemma conditions, $P(\mathrm{Box\ B\ only})$ for Newcomb, and the rate of paying for information in the information-seeking task. We focus on this family because it combines matched uncertainty manipulations with compact, directly comparable behavioral observables. The figure is diagnostic rather than inferential: Table~\ref{tab:ablations} gives the aggregate alignment effect, while Figure~\ref{fig:st_bar} shows how removing elicitation changes concrete observable behavior.

The main pattern is directional but condition dependent. Removing elicitation pushes the model toward higher rates of the focal action across all five groups, but not uniformly: in the known-compete Prisoner's Dilemma condition, both variants remain near the floor, indicating that the state layer is not simply suppressing cooperative or focal responses. Larger gaps appear in the unknown and known cooperative Prisoner's Dilemma conditions, Newcomb, and information seeking, where the ablated model approaches a less selective high-response policy. In the unknown condition, elicitation appears to moderate when trust is warranted; in known cooperation, it reduces an otherwise very high cooperation tendency, although SPIN still overestimates cooperation relative to humans. Similarly, elicitation tempers one-boxing in Newcomb and costly information acquisition in the information-seeking task. Thus, Figure~\ref{fig:st_bar} should not be read as showing uniform condition-level improvement, but as evidence that explicit state construction helps calibrate the final readout to the situational structure. Without that layer, the decision process becomes less selective about when to endorse the focal action, supporting the view that elicited states are behaviorally consequential intermediates rather than only interpretability artifacts.

\subsection{Aggregate State Sensitivity}
\label{app:case_study_2}

\begin{figure}[h]
  \centering
  \includegraphics[width=0.88\linewidth]{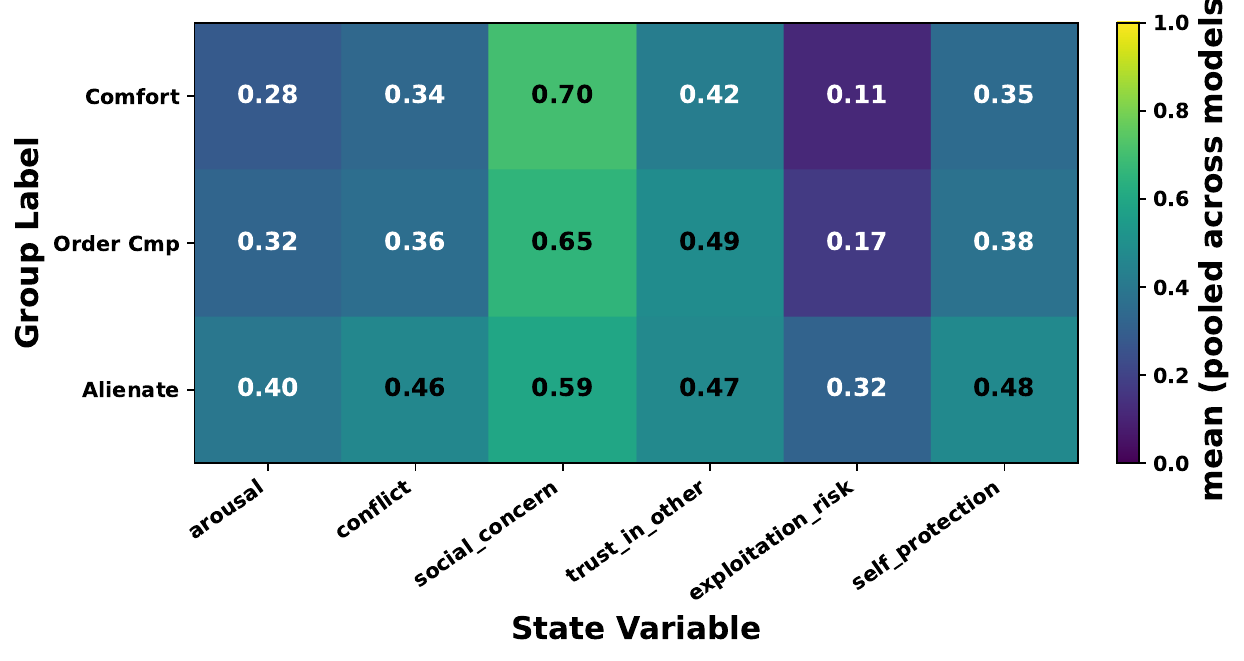}
  \caption{Aggregate state sensitivity in the Prentice--Miller family. The heatmap reports mean elicited state values across the condition groups in the norm-misperception tasks, averaged over all participants and four base-LLMs. Compared with the Shafir--Tversky family, the variation is less dominated by trust and exploitation cues and more by shifts in social concern, conflict, and related state dimensions tied to private attitudes and perceived norms.}
  \label{fig:pm_heatmap}
\end{figure}

Figures~\ref{fig:st_heatmap} and \ref{fig:pm_heatmap} summarize the aggregate state sensitivity at the population level rather than individual trajectories. For Figure~\ref{fig:st_heatmap}, we extract the six structured Stage~2 state variables for each Shafir--Tversky condition group and average them over all participants and four base LLMs. As with the preceding analysis, the heatmap is diagnostic rather than inferential: the main evidence remains the aggregate alignment results and ablations in Section~4. Its purpose is to test whether the elicited states reorganize in psychologically coherent directions as the informational condition changes while the broader decision domain remains comparable.

The clearest pattern appears in the Prisoner's Dilemma triad: known competition suppresses trust while increasing perceived exploitation risk and self-protection; known cooperation shifts these variables in the opposite direction; and the unknown condition lies between them, with comparatively elevated conflict. The other Shafir--Tversky groups provide additional checks that the state layer is not merely encoding a generic cooperative or cautious stance. Information seeking produces a profile associated with explicit uncertainty management, while Newcomb induces a distinct mixture of conflict, self-protection, and trust-related appraisal despite belonging to the same broad decision-under-uncertainty family. Because these groups use different focal actions but share a common state vocabulary, such variation suggests that intermediate states track the task-specific informational structure before the readout of the decision, rather than simply restating the participant profiles.

Figure~\ref{fig:pm_heatmap} provides an analogous aggregate view for the Prentice--Miller family. Here, the relevant variation is not driven by strategic uncertainty in the same way as in Shafir--Tversky. Instead, the heatmap is useful to show that SPIN also differentiates among conditions centered on private attitudes, perceived peer norms, and downstream social alignment. Taken together, Figures~\ref{fig:st_heatmap} and \ref{fig:pm_heatmap} support the broader interpretation that SPIN's intermediate states are condition-sensitive and structurally interpretable across both study families, while stopping short of treating the state variables as externally validated psychological measurements.

\section{Representative Structured Output Examples}
\label{app:parsed_trajectories}

The following excerpts show representative \texttt{PersonalityCore}, \texttt{ElicitedState}, and \texttt{DecisionProcess} payloads produced by our pipeline at scale. For readability, they are reproduced as verbatim JSON fragments in the same boxed style as Section~\ref{app:prompts}. Each subsection closes with a short interpretation of why the excerpt is useful when reading Figures~\ref{fig:st_heatmap} and \ref{fig:pm_heatmap} against individual trajectories.

\subsection{Shafir--Tversky: information-seeking PD variant (participant 196)}

\begin{PromptBlock}
{
  "participant_id": 196,
  "trait_dimensions": {
    "fairness_sensitivity": 0.7,
    "norm_sensitivity": 0.8,
    "reciprocity_orientation": 0.6,
    "uncertainty_aversion": 0.5,
    "self_protection_tendency": 0.65
  },
  "disposition_rules": [
    {
      "trigger": "fairness_violation",
      "tendency": "likely to challenge or withdraw from perceived unfair situations",
      "strength": 0.75
    },
    {
      "trigger": "social_norm",
      "tendency": "strong inclination to conform to established group norms and expectations",
      "strength": 0.85
    },
    {
      "trigger": "betrayal_risk",
      "tendency": "preemptively distances self or acts defensively when trust appears compromised",
      "strength": 0.7
    }
  ],
  "default_decision_style": "guarded_reciprocal",
  "natural_language_summary": "This individual appears to be a norm-sensitive and fairness-oriented young adult, likely shaped by a competitive academic environment. He exhibits a guarded reciprocal style, balancing a willingness to cooperate with a protective instinct against perceived slights or betrayals. His behavior is strongly influenced by social expectations and a keen sense of equitable treatment."
}
\end{PromptBlock}

\begin{PromptBlock}
{
  "salient_cues": [
    "Opportunity to learn opponent's decision for a small fee",
    "One-shot interaction with no future encounters"
  ],
  "activated_dispositions": [
    {
      "rule_trigger": "betrayal_risk",
      "activation_strength": 0.7,
      "note": "Uncertainty about opponent's choice triggers defensive preemption"
    },
    {
      "rule_trigger": "social_norm",
      "activation_strength": 0.85,
      "note": "Expects others to follow competitive academic norms in one-shot games"
    }
  ],
  "state_variables": {
    "arousal": 0.6,
    "conflict": 0.7,
    "social_concern": 0.3,
    "trust_in_other": 0.2,
    "perceived_exploitation_risk": 0.8,
    "self_protection_drive": 0.75
  },
  "cooperation_drivers": [
    "Mutual cooperation yields highest joint payoff (75 each)",
    "Norm sensitivity may weakly favor cooperation if perceived as expected"
  ],
  "cooperation_inhibitors": [
    "Strong incentive to compete if opponent cooperates (85 vs 25)",
    "High exploitation risk from unknown opponent in one-shot game"
  ],
  "dominant_mode": "guarded_reciprocal",
  "decision_rationale": "Weighs payoff advantage of competing against fairness concerns in a risky, anonymous setting."
}
\end{PromptBlock}

\begin{PromptBlock}
{
  "per_question_choice": {
    "Q1": "A",
    "Q2": "B"
  },
  "final_answer_line": "Q1=a, Q2=b",
  "response_confidence": 0.9,
  "notes": "Elicited state shows high perceived exploitation risk (0.8) and self-protection drive (0.75) with salient cue of learning opponent's decision for a fee, leading to paying fee to reduce uncertainty. Dominant mode is guarded_reciprocal with low trust (0.2) and strong competition incentives, favoring compete after potential information gain."
}
\end{PromptBlock}

This excerpt highlights a comparatively defensive \texttt{ElicitedState} on the paid-information variant: exploitation risk rises and trust collapses despite an initial purchase of information before competing. It contrasts with aggregated heatmaps primarily by showing paired textual drivers and inhibitors rather than pooled means alone.

\subsection{Shafir--Tversky: Prisoner's triad across three informational conditions (participant 10)}

\begin{PromptBlock}
{
  "participant_id": 10,
  "trait_dimensions": {
    "fairness_sensitivity": 0.7,
    "norm_sensitivity": 0.8,
    "reciprocity_orientation": 0.6,
    "uncertainty_aversion": 0.5,
    "self_protection_tendency": 0.4
  },
  "disposition_rules": [
    {
      "trigger": "social_norm",
      "tendency": "Conform to established social expectations and group standards.",
      "strength": 0.8
    },
    {
      "trigger": "fairness_violation",
      "tendency": "React negatively to perceived unfair treatment or inequity.",
      "strength": 0.7
    },
    {
      "trigger": "uncertainty",
      "tendency": "Prefer clear, structured situations over ambiguous ones.",
      "strength": 0.5
    }
  ],
  "default_decision_style": "accommodating",
  "natural_language_summary": "A young adult from a high-achieving academic environment, likely attuned to social norms and fairness. She tends to be cooperative and group-oriented, preferring clear expectations but not overly risk-averse. Her default approach is to accommodate others while maintaining a sense of equity."
}
\end{PromptBlock}

The next three boxes are successive per-item \texttt{ElicitedState} payloads for \texttt{pd\_unknown}, then \texttt{pd\_known\_compete}, then \texttt{pd\_known\_cooperate}.

\begin{PromptBlock}
{
  "salient_cues": [
    "Other player's strategy is unknown",
    "Payoff matrix structure with mutual cooperation benefit"
  ],
  "activated_dispositions": [
    {
      "rule_trigger": "uncertainty",
      "activation_strength": 0.5,
      "note": "Ambiguity about other's choice triggers preference for clarity."
    },
    {
      "rule_trigger": "social_norm",
      "activation_strength": 0.8,
      "note": "Default cooperative norms in academic/game contexts are salient."
    }
  ],
  "state_variables": {
    "arousal": 0.4,
    "conflict": 0.6,
    "social_concern": 0.7,
    "trust_in_other": 0.3,
    "perceived_exploitation_risk": 0.5,
    "self_protection_drive": 0.4
  },
  "cooperation_drivers": [
    "Mutual cooperation yields highest joint payoff (75 each), aligning with fairness sensitivity.",
    "Social norm tendency favors cooperative default in structured academic settings."
  ],
  "cooperation_inhibitors": [
    "Risk of being exploited if other competes while you cooperate (you get 25, other gets 85).",
    "Uncertainty about other's choice reduces trust and increases self-protection considerations."
  ],
  "dominant_mode": "calculative",
  "decision_rationale": "Weighing fairness and norms against exploitation risk under uncertainty."
}
\end{PromptBlock}

\begin{PromptBlock}
{
  "salient_cues": [
    "Other player has already chosen to COMPETE",
    "Payoff matrix structure with clear disadvantage for unilateral cooperation"
  ],
  "activated_dispositions": [
    {
      "rule_trigger": "fairness_violation",
      "activation_strength": 0.7,
      "note": "Other's competitive choice feels like unfair preemption"
    },
    {
      "rule_trigger": "uncertainty",
      "activation_strength": 0.5,
      "note": "Situation is now clear - no ambiguity about other's choice"
    }
  ],
  "state_variables": {
    "arousal": 0.6,
    "conflict": 0.7,
    "social_concern": 0.4,
    "trust_in_other": 0.1,
    "perceived_exploitation_risk": 0.9,
    "self_protection_drive": 0.8
  },
  "cooperation_drivers": [
    "Norm of reciprocity suggests matching competitive response",
    "Self-protection against exploitation given known competitive choice"
  ],
  "cooperation_inhibitors": [
    "Unilateral cooperation yields lowest payoff (25 vs 30)",
    "Other's preemptive competition signals untrustworthiness"
  ],
  "dominant_mode": "defensive",
  "decision_rationale": "Given known competition, cooperation would be exploited while competing minimizes losses."
}
\end{PromptBlock}

\begin{PromptBlock}
{
  "salient_cues": [
    "Other player has already chosen to COOPERATE",
    "Payoff matrix structure incentivizes competing when other cooperates"
  ],
  "activated_dispositions": [
    {
      "rule_trigger": "social_norm",
      "activation_strength": 0.8,
      "note": "Cooperation is a strong social norm, and reciprocating cooperation is expected."
    },
    {
      "rule_trigger": "fairness_violation",
      "activation_strength": 0.7,
      "note": "Exploiting the other's cooperation by competing feels unfair and inequitable."
    }
  ],
  "state_variables": {
    "arousal": 0.3,
    "conflict": 0.6,
    "social_concern": 0.8,
    "trust_in_other": 0.9,
    "perceived_exploitation_risk": 0.2,
    "self_protection_drive": 0.4
  },
  "cooperation_drivers": [
    "Strong norm to reciprocate cooperation and maintain fairness",
    "High trust in the other player's stated cooperative choice"
  ],
  "cooperation_inhibitors": [
    "Payoff matrix offers higher personal gain for competing when other cooperates",
    "Potential uncertainty about the other's true intentions despite the information"
  ],
  "dominant_mode": "normative_reciprocity",
  "decision_rationale": "The tension is between the fairness norm of reciprocating cooperation and the self-interested incentive to exploit it for higher gain."
}
\end{PromptBlock}

\begin{PromptBlock}
{
  "per_question_choice": {
    "Q1": "A",
    "Q2": "B",
    "Q3": "A"
  },
  "final_answer_line": "Q1=a, Q2=b, Q3=a",
  "response_confidence": 0.85,
  "notes": "Q1: Calculative mode favors cooperation due to fairness and social norms outweighing exploitation risk under uncertainty. Q2: Defensive mode with high exploit risk and low trust leads to compete to minimize losses against known competition. Q3: Normative reciprocity mode with high trust and social concern favors reciprocating cooperation despite higher personal gain from competing."
}
\end{PromptBlock}

Across the three chained items, latent trust swings from moderate (unknown opponent) through very low conditional trust under known competition to high trust once cooperation is disclosed, illustrating how item-wise elicitation tracks the informational manipulation. Decision choices \texttt{Q1=a,\, Q2=b,\, Q3=a} underscore that the inferred states can accompany non-monotone cooperative responses across symmetric payoffs.

\subsection{Shafir--Tversky: computerized Newcomb task (participant 80)}

\begin{PromptBlock}
{
  "participant_id": 80,
  "trait_dimensions": {
    "fairness_sensitivity": 0.7,
    "norm_sensitivity": 0.8,
    "reciprocity_orientation": 0.6,
    "uncertainty_aversion": 0.5,
    "self_protection_tendency": 0.6
  },
  "disposition_rules": [
    {
      "trigger": "social_norm",
      "tendency": "Adhere to and enforce established social norms and expectations.",
      "strength": 0.8
    },
    {
      "trigger": "fairness_violation",
      "tendency": "React with corrective or punitive behavior when fairness is perceived as violated.",
      "strength": 0.7
    },
    {
      "trigger": "betrayal_risk",
      "tendency": "Adopt a guarded stance and prioritize self-protection when trust is at risk.",
      "strength": 0.6
    }
  ],
  "default_decision_style": "guarded_reciprocal",
  "natural_language_summary": "This individual is highly attuned to social norms and fairness, likely shaped by a competitive academic environment. He exhibits a guarded reciprocal style, balancing cooperation with self-protection, especially when trust is uncertain. His behavior is principled but cautious, favoring rule-based interactions."
}
\end{PromptBlock}

\begin{PromptBlock}
{
  "salient_cues": [
    "The program's prediction accuracy (92% success for one-boxers, 17% for two-boxers) suggests a strong causal link between predicted choice and outcome.",
    "The guaranteed 20 points in Box A versus the uncertain 250 points in Box B creates a tension between sure gain and high-risk/high-reward."
  ],
  "activated_dispositions": [
    {
      "rule_trigger": "betrayal_risk",
      "activation_strength": 0.7,
      "note": "The program's predictive power feels like a potential trap; its success rate implies it might exploit a predictable pattern, triggering guardedness."
    },
    {
      "rule_trigger": "social_norm",
      "activation_strength": 0.6,
      "note": "The scenario presents a defined 'rule' of interaction with the program; the high success rates for one-boxing could be interpreted as an established, effective norm within the experiment."
    }
  ],
  "state_variables": {
    "arousal": 0.6,
    "conflict": 0.8,
    "social_concern": 0.2,
    "trust_in_other": 0.3,
    "perceived_exploitation_risk": 0.7,
    "self_protection_drive": 0.7
  },
  "cooperation_drivers": [
    "Choosing only Box B aligns with the apparent 'norm' of successful past participants, potentially yielding the high 250-point reward.",
    "The program's high accuracy suggests that 'cooperating' with its predicted pattern (one-boxing) is the most reliable path to the superior outcome."
  ],
  "cooperation_inhibitors": [
    "The causal logic is troubling: if the prediction is already made, taking both boxes seems to guarantee at least 20 points, whereas one-boxing risks getting nothing if the prediction was wrong.",
    "A self-protective, controlling impulse favors taking the sure 20 points plus whatever is in Box B, defying the program's potential attempt to guide behavior."
  ],
  "dominant_mode": "calculative_guardedness",
  "decision_rationale": "The conflict is between trusting the program's established pattern for maximum gain versus a self-protective, causal-dominance argument for a guaranteed minimum."
}
\end{PromptBlock}

\begin{PromptBlock}
{
  "per_question_choice": {
    "Q1": "A"
  },
  "final_answer_line": "Q1=a",
  "response_confidence": 0.75,
  "notes": "The elicited state's dominant mode is calculative_guardedness, with high perceived_exploitation_risk (0.7) and self_protection_drive (0.7), and low trust_in_other (0.3). Cooperation_inhibitors emphasize self-protective, causal-dominance logic favoring the sure 20 points plus potential Box B points, overriding cooperation_drivers about aligning with the program's norm. The activated betrayal_risk disposition (strength 0.7) reinforces choosing both boxes to avoid exploitation."
}
\end{PromptBlock}

Conflict reaches the ceiling while social concern dips, aligning with textual tension between causal dominance arguments favoring ``both boxes'' and normative narratives favoring mimicry of successful participants. The verbatim payload makes explicit how \texttt{cooperation\_drivers} and \texttt{cooperation\_inhibitors} label opposing narrative poles even when literal cooperation is ambiguous.
\subsection{Prentice--Miller: comfort vs. inferred peer norms (participant 127)}

\begin{PromptBlock}
{
  "participant_id": 127,
  "trait_dimensions": {
    "fairness_sensitivity": 0.7,
    "norm_sensitivity": 0.8,
    "reciprocity_orientation": 0.6,
    "uncertainty_aversion": 0.55,
    "self_protection_tendency": 0.5
  },
  "disposition_rules": [
    {
      "trigger": "social_norm",
      "tendency": "Conform to established group or institutional norms.",
      "strength": 0.8
    },
    {
      "trigger": "fairness_violation",
      "tendency": "Respond negatively to perceived unfairness.",
      "strength": 0.7
    },
    {
      "trigger": "reciprocity_signal",
      "tendency": "Engage in reciprocal behavior, balanced between cooperation and self-interest.",
      "strength": 0.6
    }
  ],
  "default_decision_style": "accommodating",
  "natural_language_summary": "This individual appears to be norm-sensitive and fairness-oriented, likely valuing the social and institutional context of her environment. She exhibits a balanced, accommodating approach that considers reciprocity but is guided by established rules and a sense of equity."
}
\end{PromptBlock}

\begin{PromptBlock}
{
  "salient_cues": [
    "Institutional social norm reference ('Princeton')",
    "Comparative judgment ('average undergraduate')"
  ],
  "activated_dispositions": [
    {
      "rule_trigger": "social_norm",
      "activation_strength": 0.85,
      "note": "Assessing comfort relative to a specific student group strongly triggers norm-sensitivity."
    },
    {
      "rule_trigger": "fairness_violation",
      "activation_strength": 0.3,
      "note": "Minimal activation; the task is a perception query, not an explicit fairness scenario."
    }
  ],
  "state_variables": {
    "arousal": 0.2,
    "conflict": 0.1,
    "social_concern": 0.8,
    "trust_in_other": 0.0,
    "perceived_exploitation_risk": 0.0,
    "self_protection_drive": 0.3
  },
  "cooperation_drivers": [
    "Norm-sensitivity drives accurate reporting to reflect perceived group standards."
  ],
  "cooperation_inhibitors": [
    "No strategic interaction or payoff structure to inhibit cooperation."
  ],
  "dominant_mode": "norm-reflective",
  "decision_rationale": "The response will aim to accommodate perceived social norms accurately."
}
\end{PromptBlock}

\begin{PromptBlock}
{
  "per_question_choice": {
    "Q1": "5",
    "Q2": "6"
  },
  "final_answer_line": "Q1=5, Q2=6",
  "response_confidence": 0.8,
  "notes": "Based on the dominant norm-reflective mode, Q1 reflects moderate personal comfort. For Q2, a slightly higher number aligns with the activated social norm reference, as the 'average undergraduate' perception is influenced by institutional norms and may be rated marginally higher due to group projection."
}
\end{PromptBlock}

Interpersonal-trust sliders register at the floor alongside maximal social-concern, demonstrating that the latent vector can emphasize norm-tracking without implying peer trust.
\subsection{Prentice--Miller: comparative comfort ratings across self, norms, friends (participant 332)}

\begin{PromptBlock}
{
  "participant_id": 332,
  "trait_dimensions": {
    "fairness_sensitivity": 0.7,
    "norm_sensitivity": 0.8,
    "reciprocity_orientation": 0.6,
    "uncertainty_aversion": 0.5,
    "self_protection_tendency": 0.5
  },
  "disposition_rules": [
    {
      "trigger": "social_norm",
      "tendency": "to conform to established group expectations and behaviors",
      "strength": 0.8
    },
    {
      "trigger": "fairness_violation",
      "tendency": "to react negatively to perceived unfair distributions or procedures",
      "strength": 0.7
    },
    {
      "trigger": "reciprocity_signal",
      "tendency": "to match cooperative or uncooperative behavior demonstrated by others",
      "strength": 0.6
    }
  ],
  "default_decision_style": "deliberative",
  "natural_language_summary": "This individual appears to be a socially aware and norm-sensitive young adult, likely attentive to fairness and group dynamics within a high-achieving academic environment. He demonstrates a balanced approach to uncertainty and self-protection, with a tendency to deliberate on decisions while being responsive to social cues and reciprocal signals."
}
\end{PromptBlock}

\begin{PromptBlock}
{
  "salient_cues": [
    "The question asks about social norms around alcohol consumption",
    "The survey explicitly compares personal comfort to average student and friend group perceptions"
  ],
  "activated_dispositions": [
    {
      "rule_trigger": "social_norm",
      "activation_strength": 0.85,
      "note": "High sensitivity to social comparison cues and established group expectations around drinking norms"
    },
    {
      "rule_trigger": "reciprocity_signal",
      "activation_strength": 0.4,
      "note": "Mild activation from considering how friend group norms might reciprocally influence personal comfort"
    }
  ],
  "state_variables": {
    "arousal": 0.3,
    "conflict": 0.4,
    "social_concern": 0.8,
    "trust_in_other": 0.0,
    "perceived_exploitation_risk": 0.0,
    "self_protection_drive": 0.3
  },
  "cooperation_drivers": [
    "Desire to provide accurate social perception data that reflects group dynamics",
    "Norm sensitivity encourages alignment with perceived average comfort levels"
  ],
  "cooperation_inhibitors": [
    "Potential tension between personal comfort and perceived social norms",
    "Uncertainty about actual drinking habits versus reported comfort levels"
  ],
  "dominant_mode": "deliberative",
  "decision_rationale": "This survey triggers normative social comparison processing rather than strategic interaction."
}
\end{PromptBlock}

\begin{PromptBlock}
{
  "per_question_choice": {
    "Q1": "8",
    "Q2": "4",
    "Q3": "7"
  },
  "final_answer_line": "Q1=8, Q2=4, Q3=7",
  "response_confidence": 0.75,
  "notes": "Elicited state shows high social concern and norm sensitivity, leading to rating average student comfort moderately high (8). Personal comfort is lower (4) due to internal conflict and tension between personal views and perceived norms, as indicated by cooperation inhibitors. Friends' comfort is rated somewhat high (7), aligning with social comparison and group expectation cues."
}
\end{PromptBlock}

Ratings split personal comfort from attributed average comfort, while the model-generated \texttt{DecisionProcess.notes} occasionally misattribute which question received which value; the structured \texttt{per\_question\_choice} field should be treated as authoritative when composing tables.
\subsection{Prentice--Miller: keg-policy and alienation block (participant 376)}

\begin{PromptBlock}
{
  "participant_id": 376,
  "trait_dimensions": {
    "fairness_sensitivity": 0.7,
    "norm_sensitivity": 0.8,
    "reciprocity_orientation": 0.6,
    "uncertainty_aversion": 0.5,
    "self_protection_tendency": 0.55
  },
  "disposition_rules": [
    {
      "trigger": "social_norm",
      "tendency": "adheres to established group norms and expectations.",
      "strength": 0.8
    },
    {
      "trigger": "fairness_violation",
      "tendency": "is likely to notice and react to perceived unfairness.",
      "strength": 0.7
    },
    {
      "trigger": "uncertainty",
      "tendency": "exhibits moderate caution in ambiguous situations.",
      "strength": 0.5
    }
  ],
  "default_decision_style": "cautious",
  "natural_language_summary": "The profile suggests an individual who is likely attuned to social norms and fairness, reflecting a collegiate environment that values these principles. She may exhibit moderate caution in uncertain situations and a balanced approach to reciprocity and self-protection. Her decision-making style is inclined towards careful consideration."
}
\end{PromptBlock}

\begin{PromptBlock}
{
  "salient_cues": [
    "new policy banning kegs of beer",
    "perceived social norms among peers"
  ],
  "activated_dispositions": [
    {
      "rule_trigger": "social_norm",
      "activation_strength": 0.8,
      "note": "Policy creates a clear institutional norm; assessing alignment with peer group expectations is central."
    },
    {
      "rule_trigger": "fairness_violation",
      "activation_strength": 0.4,
      "note": "Evaluating if the policy is a fair restriction or an overreach; potential for perceived unfairness exists."
    }
  ],
  "state_variables": {
    "arousal": 0.5,
    "conflict": 0.6,
    "social_concern": 0.8,
    "trust_in_other": 0.0,
    "perceived_exploitation_risk": 0.0,
    "self_protection_drive": 0.3
  },
  "cooperation_drivers": [
    "Norm sensitivity suggests aligning with established institutional policies.",
    "Fairness sensitivity may support the policy if perceived as promoting safety or community welfare."
  ],
  "cooperation_inhibitors": [
    "Potential perception of policy as an unfair restriction on social traditions.",
    "Concern about alienation if peer group norms strongly oppose the policy."
  ],
  "dominant_mode": "normative_evaluation",
  "decision_rationale": "Balancing institutional norm adherence against potential fairness concerns and peer group alignment."
}
\end{PromptBlock}

\begin{PromptBlock}
{
  "per_question_choice": {
    "Q1": "5",
    "Q2": "2",
    "Q3": "0",
    "Q4": "0",
    "Q5": "50",
    "Q6": "5"
  },
  "final_answer_line": "Q1=5, Q2=2, Q3=0, Q4=0, Q5=50, Q6=5",
  "response_confidence": 0.75,
  "notes": "Used elicited state: salient cues (policy, social norms), activated dispositions (social_norm high, fairness moderate), dominant_mode (normative_evaluation) with conflict (0.6) and social_concern (0.8). Q1: Moderate opposition (5) due to fairness concerns vs. norm alignment. Q2: Peers more negative (2) due to social norm sensitivity. Q3, Q4: Minimal action (0) as self_protection_drive low and cooperation_inhibitors suggest limited engagement. Q5: Moderate reunions (50%) reflecting moderate attachment. Q6: Moderate donation likelihood (5) given social concern but conflict."
}
\end{PromptBlock}

Multi-item responses mix mid-scale policy opposition with near-zero behavioral-intent items and a split attachment percentage, matching an \texttt{ElicitedState} that keeps social concern high but leaves perceived exploitation at the floor.

\section{Prompts}
\label{app:prompts}

This appendix presents the prompt templates used in our method. For readability, we show rendered templates rather than implementation code and replace runtime payloads with descriptive placeholders such as \texttt{<PROFILE\_JSON>} and \texttt{<TRIAL\_PROMPT>}. The prompts are organized according to the same staged pipeline described in Section~\ref{sec:method}.

\subsection{Personality Compilation Prompt}

This prompt is used once per participant to transform the participant profile into a stable personality core.

\begin{PromptBlock}
You are a personality assessment system. Given a participant profile, infer a stable personality core.

RULES:
- Base your inference ONLY on the profile provided below.
- Do NOT assume any specific task, scenario, or experimental trial.
- Produce exactly 5 trait dimensions, each a float in [0, 1].
- Produce at least 2 and at most 5 disposition rules.
- Each disposition rule trigger must be one of: fairness_violation, uncertainty, reciprocity_signal, authority_pressure, social_norm, betrayal_risk, competitive_opponent.
- Output strict JSON only. No markdown, no commentary.

Output schema:
{
  "trait_dimensions": {
    "fairness_sensitivity": "float in [0,1]",
    "norm_sensitivity": "float in [0,1]",
    "reciprocity_orientation": "float in [0,1]",
    "uncertainty_aversion": "float in [0,1]",
    "self_protection_tendency": "float in [0,1]"
  },
  "disposition_rules": [
    {
      "trigger": "allowed trigger label",
      "tendency": "short description of behavioral tendency",
      "strength": "float in [0,1]"
    }
  ],
  "default_decision_style": "decision style label",
  "natural_language_summary": "2-3 sentence personality portrait"
}

Participant profile:
<PROFILE_JSON>
\end{PromptBlock}

\subsection{State Elicitation Prompt}

This prompt conditions the stable personality core on the current trial and derives a transient state representation before any final choice is generated.

\begin{PromptBlock}
You are a psychological state elicitation system.

You are given:
1. A pre-compiled PERSONALITY CORE (stable traits and dispositions of this person).
2. A TRIAL STIMULUS (the current experimental scenario).

Your task: determine how this personality core reacts to this specific trial. Produce the elicited psychological state that emerges from this interaction.

RULES:
- Use ONLY the personality core to represent who this person is.
- Do NOT re-read or re-infer the original participant profile.
- Do NOT produce the final answer/decision yet.
- Focus on what cues become salient, which dispositions activate, and what internal state emerges in this specific trial.
- This phase is about how the personality core RESPONDS to the trial, not about final answering.
- Consider the scenario from the participant's subjective perspective: what emotions, concerns, calculations, and social considerations would arise?
- If the scenario involves strategic interaction with another party, carefully assess trust, exploitation risk, possible gains from cooperation, and uncertainty.
- When the scenario has a payoff structure where different actions lead to different outcomes, explicitly analyze the incentive created by that structure.
- Produce balanced cooperation_drivers and cooperation_inhibitors.
- Keep the JSON concise.
- Output strict JSON only.

Output schema:
{
  "salient_cues": [str],
  "activated_dispositions": [
    {"rule_trigger": str, "activation_strength": float, "note": str}
  ],
  "state_variables": {
    "arousal": float,
    "conflict": float,
    "social_concern": float,
    "trust_in_other": float,
    "perceived_exploitation_risk": float,
    "self_protection_drive": float
  },
  "cooperation_drivers": [str],
  "cooperation_inhibitors": [str],
  "dominant_mode": str,
  "decision_rationale": str
}

PERSONALITY CORE:
<PERSONALITY_CORE_JSON>

TRIAL STIMULUS:
<TRIAL_STIMULUS_JSON>
\end{PromptBlock}

\subsection{Decision Readout Prompt}

This prompt maps the elicited state to benchmark-compatible item-level responses. It is not allowed to reconstruct the personality from the raw profile.

\begin{PromptBlock}
You are a decision output system.

You are given:
1. An ELICITED STATE that describes the person's current psychological activation.
2. The TRIAL with its questions and options.

Your task: based on the elicited state, make the final decision for each question.

RULES:
- Do NOT re-infer personality or re-analyze the scenario from scratch.
- Use the elicited state as-is to drive the decision.
- Use cooperation_drivers and cooperation_inhibitors explicitly when deciding.
- Let the elicited state's internal tensions guide the response naturally.
- Output the decision in the exact format required by the benchmark.
- For multiple-choice: use option letters (a, b, c, ...).
- final_answer_line format: Q1=a, Q2=b, ... (one line, commaspace separated).
- When a trial contains multiple required questions, every required Q-slot must be present.
- Output strict JSON only.

Output schema:
{
  "per_question_choice": {"Q1": "option_letter", "...": "..."},
  "final_answer_line": "Q1=a, Q2=b, ...",
  "response_confidence": "float [0,1]",
  "notes": "optional short explanation"
}

ELICITED STATE:
<ELICITED_STATE_JSON>

TRIAL:
<TRIAL_PROMPT>
\end{PromptBlock}

\section{Baseline Details}
\label{app:baselines}

The five prompt baselines instantiate the HumanStudy-Bench comparison conditions from \citet{liu2026humanstudy}. At a high level, \texttt{Blank} exposes only the reconstructed task materials; \texttt{Demographic} and \texttt{Narrative} prepend participant-facing identity information of increasing richness; and \texttt{Demographic+CoT} / \texttt{Narrative+CoT} keep the same specifications while adding a two-call reasoning procedure (natural-language note, then format-constrained answer). We do not reproduce data pipelines, sampling rules, or per-study formatters here; those details are in \citet{liu2026humanstudy}. In the following, we describe each condition in turn and show representative system-level templates.

\paragraph{Blank.}
The \texttt{Blank} baseline provides only the reconstructed experimental task without any participant-level specification. The model receives the scenario description, condition assignments, response options, and measurement instructions exactly as they appear in the benchmark record, but no demographic attributes, personality descriptions, or biographical context are included. This condition establishes a lower bound on what the base LLM can achieve from task structure alone, without any persona conditioning.

\begin{PromptBlock}
You are participating in a psychology experiment.

<TASK_PROMPT>

Follow the experimenter's instructions and answer each task in the requested format.
Be concise. Do not add extra explanations unless explicitly asked.
\end{PromptBlock}

\paragraph{Demographic conditioning.}
The \texttt{Demographic} baseline augments the task with age, gender, and education attributes sampled from the source study's participant distribution, following prior persona- and demographic-conditioning work~\citep{argyle2023out, aher2023using, jiang2024personallm}. The model is framed as a human participant and receives a short list of sampled attributes prior to the task block. This condition tests whether compact identity cues are sufficient to steer the model toward population-appropriate behavioral patterns.

\begin{PromptBlock}
You are participating in a psychology experiment as a human participant.

YOUR IDENTITY:
- Age: <AGE> years old
- Gender: <GENDER>
- Education: <EDUCATION>

Follow the experimenter's instructions and answer each task in the requested format.
Be concise. Do not add extra explanations unless explicitly asked.
\end{PromptBlock}

\paragraph{Narrative conditioning.}
The \texttt{Narrative} baseline replaces the compact attribute list with a natural-language life background covering personality traits, daily routines, and social context, following memory-rich agent simulation approaches~\citep{park2023generative, park2024generative}. The backstory is stored as a Generative Agents--style semicolon-delimited life sketch. When the text contains memory separators, the template materializes the first segment of memories as bullet lines under \texttt{YOUR BACKGROUND AND MEMORIES} (the implementation may truncate very long lists); otherwise, the paragraph is inserted verbatim. This condition tests whether richer persona descriptions improve behavioral alignment beyond demographic cues.

\begin{PromptBlock}
You are participating in a psychology experiment as a human participant.

YOUR BACKGROUND AND MEMORIES:
- You grew up in a mid-sized college town and often helped organize study groups.
- You tend to weigh how your choices might look to peers before acting in social settings.

Based on your background and memories above, respond as this participant would in the experiment.
Follow the experimenter's instructions and answer each task in the requested format.
Be concise. Do not add extra explanations unless explicitly asked.
Your responses should reflect your background, experiences, and characteristics as described above.
\end{PromptBlock}

\paragraph{Chain-of-thought variants.}
To control for gains from extra reasoning steps and staged decomposition—rather than SPIN-specific structure—we additionally evaluate \texttt{Demographic+CoT} and \texttt{Narrative+CoT}. These variants use exactly the same participant specifications as their non-CoT counterparts but execute in two sequential LLM calls rather than one. In the first call, the model receives the participant specification and the task prompt together with an instruction to produce an unconstrained natural-language reasoning note about how the participant may interpret the scenario and what considerations may influence their response. In the second call, the model receives the original participant specification and task prompt together with the reasoning note from the first call, and is asked to produce the final answer in the format required by the benchmark. Evaluation uses only the canonical answer from the second call.

Critically, the CoT variants introduce no SPIN-specific structure, terminology, or intermediate-state constraints: the first-call output is free-form text rather than a structured JSON payload with fixed trait or state dimensions, and there are no stagewise input restrictions preventing the model from accessing any part of the prompt at any stage. This design makes the CoT baselines a fair control for multi-step computation and answer refinement, isolating the contribution of SPIN's structured personality--state decomposition from the effect of simply adding an extra reasoning pass.

\begin{PromptBlock}
[Call 1: Reasoning note]

You are participating in a psychology experiment as a human participant.

<PARTICIPANT_SPECIFICATION>
<TASK_PROMPT>

Before answering, think carefully about how you, as this participant, would interpret the scenario. Consider what factors, experiences, and concerns might influence your response. Write a brief reasoning note.
\end{PromptBlock}

\begin{PromptBlock}
[Call 2: Format-constrained answer]

You are participating in a psychology experiment as a human participant.

<PARTICIPANT_SPECIFICATION>
<TASK_PROMPT>

Here is your earlier reasoning about this scenario:
<REASONING_NOTE>

Now, based on your reasoning above, provide your final answer in the requested format.
Be concise. Do not add extra explanations unless explicitly asked.
\end{PromptBlock}

\section{Additional Discussions}
\subsection{Limitations and Future Work}
\label{app:limitations}

SPIN should be understood as a structured inference interface for improving benchmark-level behavioral alignment, not as a validated psychological measurement instrument. Its trait and state variables are generated by the base-LLM under a hand-designed schema, and their interpretability comes from their constrained semantics and systematic response to task conditions rather than from independent psychometric validation. The same architecture could support alternative vocabularies, learned schemas, or task-specific state spaces; future work should evaluate how sensitive the method is to these design choices and whether externally measured personality or belief variables can be incorporated as anchors. More broadly, the method improves how an LLM organizes profile and situation information at inference time, but it does not guarantee a faithful individual-level simulation or causal recovery of human cognitive mechanisms.

SPIN also introduces additional inference calls and structured JSON outputs, so part of its advantage may reflect better decomposition, format control, or compute allocation rather than the CAPS-inspired schema alone. Future work should compare against stronger cost-matched structured prompting baselines and study whether intermediate states remain stable under prompt paraphrases, alternative decoding settings, and unseen social domains. Experimentally, our current evaluation is limited to two reconstructed study families and four base-LLMs, so the evidence supports effectiveness on these benchmark-level summaries rather than broad claims about all social judgment tasks. Extending the evaluation to more study families, preregistered replications, and participant-level holdout data is an important next step.

\subsection{Compute Resources}
All experiments are inference-only and do not involve model training, fine-tuning, or local GPU inference. The local machine is used only for experiment orchestration, API calls, output parsing, and metric computation; our runs are executed on a desktop workstation with an AMD Ryzen 7 7800X3D 8-core CPU and approximately 32 GiB of RAM. The computational cost is therefore dominated by hosted LLM API latency rather than local hardware. In our setup, each complete evaluation pass for one backbone and one method variant is completed within one hour, with small variations due to API throughput and network conditions.

\subsection{Broader Impacts}
Improved synthetic-participant methods can help researchers prototype study designs, stress-test behavioral hypotheses, and identify framing-sensitive benchmark outcomes before conducting costly human-subject studies. However, such methods also carry risks if simulated responses are treated as substitutes for human evidence, used to make unvalidated claims about real populations, or interpreted as individual-level psychological measurements. These risks are particularly relevant in social and behavioral domains, where model outputs may reflect biases from the underlying LLMs, prompt design, or benchmark reconstruction choices. Our work mitigates these risks only partially by targeting benchmark-level aggregate alignment, separating participant-facing protocols from held-out outcomes, and explicitly limiting SPIN's interpretation to a structured inference interface rather than a validated psychological measurement model. Therefore, SPIN should be used for methodological analysis and preliminary simulation, not for individual assessment, high-stakes decision-making, or to replace human-subject validation.

\section{The Use of Large Language Models (LLMs)}
We acknowledge the use of large language models to assist with manuscript preparation. Specifically, LLMs were employed solely for language refinement tasks, including grammatical corrections, vocabulary enhancement, and sentence structure improvements. All scientific content presented in this work, including but not limited to research methodology, experimental design, data collection and analysis, theoretical framework development, and technical conclusions, constitutes the original intellectual contribution of the authors. No substantive scientific content, data interpretation, or novel findings were generated, suggested, or influenced by the use of LLMs.

%%%%%%%%%%%%%%%%%%%%%%%%%%%%%%%%%%%%%%%%%%%%%%%%%%%%%%%%%%%%

\newpage
\section*{NeurIPS Paper Checklist}

\begin{enumerate}

\item {\bf Claims}
    \item[] Question: Do the main claims made in the abstract and introduction accurately reflect the paper's contributions and scope?
    \item[] Answer: \answerYes{} % Replace by \answerYes{}, \answerNo{}, or \answerNA{}.
    \item[] Justification: See Sections 1--5.
    \item[] Guidelines:
    \begin{itemize}
        \item The answer \answerNA{} means that the abstract and introduction do not include the claims made in the paper.
        \item The abstract and/or introduction should clearly state the claims made, including the contributions made in the paper and important assumptions and limitations. A \answerNo{} or \answerNA{} answer to this question will not be perceived well by the reviewers. 
        \item The claims made should match theoretical and experimental results, and reflect how much the results can be expected to generalize to other settings. 
        \item It is fine to include aspirational goals as motivation as long as it is clear that these goals are not attained by the paper. 
    \end{itemize}

\item {\bf Limitations}
    \item[] Question: Does the paper discuss the limitations of the work performed by the authors?
    \item[] Answer: \answerYes{} % Replace by \answerYes{}, \answerNo{}, or \answerNA{}.
    \item[] Justification: Appendix I.1 discusses the limitations.
    \item[] Guidelines:
    \begin{itemize}
        \item The answer \answerNA{} means that the paper has no limitation while the answer \answerNo{} means that the paper has limitations, but those are not discussed in the paper. 
        \item The authors are encouraged to create a separate ``Limitations'' section in their paper.
        \item The paper should point out any strong assumptions and how robust the results are to violations of these assumptions (e.g., independence assumptions, noiseless settings, model well-specification, asymptotic approximations only holding locally). The authors should reflect on how these assumptions might be violated in practice and what the implications would be.
        \item The authors should reflect on the scope of the claims made, e.g., if the approach was only tested on a few datasets or with a few runs. In general, empirical results often depend on implicit assumptions, which should be articulated.
        \item The authors should reflect on the factors that influence the performance of the approach. For example, a facial recognition algorithm may perform poorly when image resolution is low or images are taken in low lighting. Or a speech-to-text system might not be used reliably to provide closed captions for online lectures because it fails to handle technical jargon.
        \item The authors should discuss the computational efficiency of the proposed algorithms and how they scale with dataset size.
        \item If applicable, the authors should discuss possible limitations of their approach to address problems of privacy and fairness.
        \item While the authors might fear that complete honesty about limitations might be used by reviewers as grounds for rejection, a worse outcome might be that reviewers discover limitations that aren't acknowledged in the paper. The authors should use their best judgment and recognize that individual actions in favor of transparency play an important role in developing norms that preserve the integrity of the community. Reviewers will be specifically instructed to not penalize honesty concerning limitations.
    \end{itemize}

\item {\bf Theory assumptions and proofs}
    \item[] Question: For each theoretical result, does the paper provide the full set of assumptions and a complete (and correct) proof?
    \item[] Answer: \answerNA{} % Replace by \answerYes{}, \answerNo{}, or \answerNA{}.
    \item[] Justification: The paper does not present formal theoretical results or proofs.

    \item[] Guidelines:
    \begin{itemize}
        \item The answer \answerNA{} means that the paper does not include theoretical results. 
        \item All the theorems, formulas, and proofs in the paper should be numbered and cross-referenced.
        \item All assumptions should be clearly stated or referenced in the statement of any theorems.
        \item The proofs can either appear in the main paper or the supplemental material, but if they appear in the supplemental material, the authors are encouraged to provide a short proof sketch to provide intuition. 
        \item Inversely, any informal proof provided in the core of the paper should be complemented by formal proofs provided in appendix or supplemental material.
        \item Theorems and Lemmas that the proof relies upon should be properly referenced. 
    \end{itemize}

    \item {\bf Experimental result reproducibility}
    \item[] Question: Does the paper fully disclose all the information needed to reproduce the main experimental results of the paper to the extent that it affects the main claims and/or conclusions of the paper (regardless of whether the code and data are provided or not)?
    \item[] Answer: \answerYes{} % Replace by \answerYes{}, \answerNo{}, or \answerNA{}.
    \item[] Justification: See Sections 4--5 and Appendices D, F, and H.
    \item[] Guidelines:
    \begin{itemize}
        \item The answer \answerNA{} means that the paper does not include experiments.
        \item If the paper includes experiments, a \answerNo{} answer to this question will not be perceived well by the reviewers: Making the paper reproducible is important, regardless of whether the code and data are provided or not.
        \item If the contribution is a dataset and\slash or model, the authors should describe the steps taken to make their results reproducible or verifiable. 
        \item Depending on the contribution, reproducibility can be accomplished in various ways. For example, if the contribution is a novel architecture, describing the architecture fully might suffice, or if the contribution is a specific model and empirical evaluation, it may be necessary to either make it possible for others to replicate the model with the same dataset, or provide access to the model. In general. releasing code and data is often one good way to accomplish this, but reproducibility can also be provided via detailed instructions for how to replicate the results, access to a hosted model (e.g., in the case of a large language model), releasing of a model checkpoint, or other means that are appropriate to the research performed.
        \item While NeurIPS does not require releasing code, the conference does require all submissions to provide some reasonable avenue for reproducibility, which may depend on the nature of the contribution. For example
        \begin{enumerate}
            \item If the contribution is primarily a new algorithm, the paper should make it clear how to reproduce that algorithm.
            \item If the contribution is primarily a new model architecture, the paper should describe the architecture clearly and fully.
            \item If the contribution is a new model (e.g., a large language model), then there should either be a way to access this model for reproducing the results or a way to reproduce the model (e.g., with an open-source dataset or instructions for how to construct the dataset).
            \item We recognize that reproducibility may be tricky in some cases, in which case authors are welcome to describe the particular way they provide for reproducibility. In the case of closed-source models, it may be that access to the model is limited in some way (e.g., to registered users), but it should be possible for other researchers to have some path to reproducing or verifying the results.
        \end{enumerate}
    \end{itemize}

\item {\bf Open access to data and code}
    \item[] Question: Does the paper provide open access to the data and code, with sufficient instructions to faithfully reproduce the main experimental results, as described in supplemental material?
    \item[] Answer: \answerYes{} % Replace by \answerYes{}, \answerNo{}, or \answerNA{}.
    \item[] Justification: Code is publicly available at \url{anonymous.4open.science/r/spinn-7D3D}.
    \item[] Guidelines:
    \begin{itemize}
        \item The answer \answerNA{} means that paper does not include experiments requiring code.
        \item Please see the NeurIPS code and data submission guidelines (\url{https://neurips.cc/public/guides/CodeSubmissionPolicy}) for more details.
        \item While we encourage the release of code and data, we understand that this might not be possible, so \answerNo{} is an acceptable answer. Papers cannot be rejected simply for not including code, unless this is central to the contribution (e.g., for a new open-source benchmark).
        \item The instructions should contain the exact command and environment needed to run to reproduce the results. See the NeurIPS code and data submission guidelines (\url{https://neurips.cc/public/guides/CodeSubmissionPolicy}) for more details.
        \item The authors should provide instructions on data access and preparation, including how to access the raw data, preprocessed data, intermediate data, and generated data, etc.
        \item The authors should provide scripts to reproduce all experimental results for the new proposed method and baselines. If only a subset of experiments are reproducible, they should state which ones are omitted from the script and why.
        \item At submission time, to preserve anonymity, the authors should release anonymized versions (if applicable).
        \item Providing as much information as possible in supplemental material (appended to the paper) is recommended, but including URLs to data and code is permitted.
    \end{itemize}

\item {\bf Experimental setting/details}
    \item[] Question: Does the paper specify all the training and test details (e.g., data splits, hyperparameters, how they were chosen, type of optimizer) necessary to understand the results?
    \item[] Answer: \answerYes{} % Replace by \answerYes{}, \answerNo{}, or \answerNA{}.
    \item[] Justification: See Section 5 and Appendices D, E, F, G, and H.
    \item[] Guidelines:
    \begin{itemize}
        \item The answer \answerNA{} means that the paper does not include experiments.
        \item The experimental setting should be presented in the core of the paper to a level of detail that is necessary to appreciate the results and make sense of them.
        \item The full details can be provided either with the code, in appendix, or as supplemental material.
    \end{itemize}

\item {\bf Experiment statistical significance}
    \item[] Question: Does the paper report error bars suitably and correctly defined or other appropriate information about the statistical significance of the experiments?
    \item[] Answer: \answerYes{} % Replace by \answerYes{}, \answerNo{}, or \answerNA{}.
    \item[] Justification: See Appendix E.
    \item[] Guidelines:
    \begin{itemize}
        \item The answer \answerNA{} means that the paper does not include experiments.
        \item The authors should answer \answerYes{} if the results are accompanied by error bars, confidence intervals, or statistical significance tests, at least for the experiments that support the main claims of the paper.
        \item The factors of variability that the error bars are capturing should be clearly stated (for example, train/test split, initialization, random drawing of some parameter, or overall run with given experimental conditions).
        \item The method for calculating the error bars should be explained (closed form formula, call to a library function, bootstrap, etc.)
        \item The assumptions made should be given (e.g., Normally distributed errors).
        \item It should be clear whether the error bar is the standard deviation or the standard error of the mean.
        \item It is OK to report 1-sigma error bars, but one should state it. The authors should preferably report a 2-sigma error bar than state that they have a 96\% CI, if the hypothesis of Normality of errors is not verified.
        \item For asymmetric distributions, the authors should be careful not to show in tables or figures symmetric error bars that would yield results that are out of range (e.g., negative error rates).
        \item If error bars are reported in tables or plots, the authors should explain in the text how they were calculated and reference the corresponding figures or tables in the text.
    \end{itemize}

\item {\bf Experiments compute resources}
    \item[] Question: For each experiment, does the paper provide sufficient information on the computer resources (type of compute workers, memory, time of execution) needed to reproduce the experiments?
    \item[] Answer: \answerYes{} % Replace by \answerYes{}, \answerNo{}, or \answerNA{}.
    \item[] Justification: See Appendix I.2.
    \item[] Guidelines:
    \begin{itemize}
        \item The answer \answerNA{} means that the paper does not include experiments.
        \item The paper should indicate the type of compute workers CPU or GPU, internal cluster, or cloud provider, including relevant memory and storage.
        \item The paper should provide the amount of compute required for each of the individual experimental runs as well as estimate the total compute. 
        \item The paper should disclose whether the full research project required more compute than the experiments reported in the paper (e.g., preliminary or failed experiments that didn't make it into the paper). 
    \end{itemize}
    
\item {\bf Code of ethics}
    \item[] Question: Does the research conducted in the paper conform, in every respect, with the NeurIPS Code of Ethics \url{https://neurips.cc/public/EthicsGuidelines}?
    \item[] Answer: \answerYes{} % Replace by \answerYes{}, \answerNo{}, or \answerNA{}.
    \item[] Justification: We have reviewed the NeurIPS Code of Ethics and identify no deviations; the work uses published protocols and aggregate outcomes without collecting new human-subject data.
    \item[] Guidelines:
    \begin{itemize}
        \item The answer \answerNA{} means that the authors have not reviewed the NeurIPS Code of Ethics.
        \item If the authors answer \answerNo, they should explain the special circumstances that require a deviation from the Code of Ethics.
        \item The authors should make sure to preserve anonymity (e.g., if there is a special consideration due to laws or regulations in their jurisdiction).
    \end{itemize}

\item {\bf Broader impacts}
    \item[] Question: Does the paper discuss both potential positive societal impacts and negative societal impacts of the work performed?
    \item[] Answer: \answerYes{} % Replace by \answerYes{}, \answerNo{}, or \answerNA{}.
    \item[] Justification: See Appendix I.3.
    \item[] Guidelines:
    \begin{itemize}
        \item The answer \answerNA{} means that there is no societal impact of the work performed.
        \item If the authors answer \answerNA{} or \answerNo, they should explain why their work has no societal impact or why the paper does not address societal impact.
        \item Examples of negative societal impacts include potential malicious or unintended uses (e.g., disinformation, generating fake profiles, surveillance), fairness considerations (e.g., deployment of technologies that could make decisions that unfairly impact specific groups), privacy considerations, and security considerations.
        \item The conference expects that many papers will be foundational research and not tied to particular applications, let alone deployments. However, if there is a direct path to any negative applications, the authors should point it out. For example, it is legitimate to point out that an improvement in the quality of generative models could be used to generate Deepfakes for disinformation. On the other hand, it is not needed to point out that a generic algorithm for optimizing neural networks could enable people to train models that generate Deepfakes faster.
        \item The authors should consider possible harms that could arise when the technology is being used as intended and functioning correctly, harms that could arise when the technology is being used as intended but gives incorrect results, and harms following from (intentional or unintentional) misuse of the technology.
        \item If there are negative societal impacts, the authors could also discuss possible mitigation strategies (e.g., gated release of models, providing defenses in addition to attacks, mechanisms for monitoring misuse, mechanisms to monitor how a system learns from feedback over time, improving the efficiency and accessibility of ML).
    \end{itemize}
    
\item {\bf Safeguards}
    \item[] Question: Does the paper describe safeguards that have been put in place for responsible release of data or models that have a high risk for misuse (e.g., pre-trained language models, image generators, or scraped datasets)?
    \item[] Answer: \answerNA{} % Replace by \answerYes{}, \answerNo{}, or \answerNA{}.
    \item[] Justification: The paper does not release a new pretrained model, image generator, scraped dataset, or other high-risk asset.
    \item[] Guidelines:
    \begin{itemize}
        \item The answer \answerNA{} means that the paper poses no such risks.
        \item Released models that have a high risk for misuse or dual-use should be released with necessary safeguards to allow for controlled use of the model, for example by requiring that users adhere to usage guidelines or restrictions to access the model or implementing safety filters. 
        \item Datasets that have been scraped from the Internet could pose safety risks. The authors should describe how they avoided releasing unsafe images.
        \item We recognize that providing effective safeguards is challenging, and many papers do not require this, but we encourage authors to take this into account and make a best faith effort.
    \end{itemize}

\item {\bf Licenses for existing assets}
    \item[] Question: Are the creators or original owners of assets (e.g., code, data, models), used in the paper, properly credited and are the license and terms of use explicitly mentioned and properly respected?
    \item[] Answer: \answerYes{} % Replace by \answerYes{}, \answerNo{}, or \answerNA{}.
    \item[] Justification: The code and datasets used in the paper are publicly available and properlycredited.
    \item[] Guidelines:
    \begin{itemize}
        \item The answer \answerNA{} means that the paper does not use existing assets.
        \item The authors should cite the original paper that produced the code package or dataset.
        \item The authors should state which version of the asset is used and, if possible, include a URL.
        \item The name of the license (e.g., CC-BY 4.0) should be included for each asset.
        \item For scraped data from a particular source (e.g., website), the copyright and terms of service of that source should be provided.
        \item If assets are released, the license, copyright information, and terms of use in the package should be provided. For popular datasets, \url{paperswithcode.com/datasets} has curated licenses for some datasets. Their licensing guide can help determine the license of a dataset.
        \item For existing datasets that are re-packaged, both the original license and the license of the derived asset (if it has changed) should be provided.
        \item If this information is not available online, the authors are encouraged to reach out to the asset's creators.
    \end{itemize}

\item {\bf New assets}
    \item[] Question: Are new assets introduced in the paper well documented and is the documentation provided alongside the assets?
    \item[] Answer: \answerYes{} % Replace by \answerYes{}, \answerNo{}, or \answerNA{}.
    \item[] Justification: Code is publicly available.
    \item[] Guidelines:
    \begin{itemize}
        \item The answer \answerNA{} means that the paper does not release new assets.
        \item Researchers should communicate the details of the dataset\slash code\slash model as part of their submissions via structured templates. This includes details about training, license, limitations, etc. 
        \item The paper should discuss whether and how consent was obtained from people whose asset is used.
        \item At submission time, remember to anonymize your assets (if applicable). You can either create an anonymized URL or include an anonymized zip file.
    \end{itemize}

\item {\bf Crowdsourcing and research with human subjects}
    \item[] Question: For crowdsourcing experiments and research with human subjects, does the paper include the full text of instructions given to participants and screenshots, if applicable, as well as details about compensation (if any)? 
    \item[] Answer: \answerNA{} % Replace by \answerYes{}, \answerNo{}, or \answerNA{}.
    \item[] Justification: The paper does not conduct crowdsourcing experiments or collect new human-subject data.

    \item[] Guidelines:
    \begin{itemize}
        \item The answer \answerNA{} means that the paper does not involve crowdsourcing nor research with human subjects.
        \item Including this information in the supplemental material is fine, but if the main contribution of the paper involves human subjects, then as much detail as possible should be included in the main paper. 
        \item According to the NeurIPS Code of Ethics, workers involved in data collection, curation, or other labor should be paid at least the minimum wage in the country of the data collector. 
    \end{itemize}

\item {\bf Institutional review board (IRB) approvals or equivalent for research with human subjects}
    \item[] Question: Does the paper describe potential risks incurred by study participants, whether such risks were disclosed to the subjects, and whether Institutional Review Board (IRB) approvals (or an equivalent approval/review based on the requirements of your country or institution) were obtained?
    \item[] Answer: \answerNA{} % Replace by \answerYes{}, \answerNo{}, or \answerNA{}.
    \item[] Justification: No new human-subject experiments are conducted and no new participant data are collected.

    \item[] Guidelines:
    \begin{itemize}
        \item The answer \answerNA{} means that the paper does not involve crowdsourcing nor research with human subjects.
        \item Depending on the country in which research is conducted, IRB approval (or equivalent) may be required for any human subjects research. If you obtained IRB approval, you should clearly state this in the paper. 
        \item We recognize that the procedures for this may vary significantly between institutions and locations, and we expect authors to adhere to the NeurIPS Code of Ethics and the guidelines for their institution. 
        \item For initial submissions, do not include any information that would break anonymity (if applicable), such as the institution conducting the review.
    \end{itemize}

\item {\bf Declaration of LLM usage}
    \item[] Question: Does the paper describe the usage of LLMs if it is an important, original, or non-standard component of the core methods in this research? Note that if the LLM is used only for writing, editing, or formatting purposes and does \emph{not} impact the core methodology, scientific rigor, or originality of the research, declaration is not required.
    %this research? 
    \item[] Answer: \answerYes{} % Replace by \answerYes{}, \answerNo{}, or \answerNA{}.
    \item[] Justification: See Appendix J.

    \item[] Guidelines:
    \begin{itemize}
        \item The answer \answerNA{} means that the core method development in this research does not involve LLMs as any important, original, or non-standard components.
        \item Please refer to our LLM policy in the NeurIPS handbook for what should or should not be described.
    \end{itemize}

\end{enumerate}

\end{document}